\documentclass{aa}  
\usepackage{natbib}

\usepackage{graphicx}
\usepackage[varg]{txfonts}
\def\Ou  {O\,{\small I}}
\def\Cu  {C\,{\small I}}

\def\Feu  {Fe\,{\small I}}
\def\Fed  {Fe\,{\small II}}
\def\Cru  {Cr\,{\small I}}
\def\Crd  {Cr\,{\small II}}

\def\Mnu  {Mn\,{\small I}}
\def\Teff  {$T_\mathrm{eff}$}

\def\loggf {$\log gf$}
\def\kms   {$\rm km s^{-1}$}
\newcommand{\xx}{\ensuremath{\mathrm{1D}_{\mathrm{LHD}}}}
\newcommand{\mD}{\ensuremath{\left\langle\mathrm{3D}\right\rangle}}

\newcommand{\cobold}{{\sf CO$^5$BOLD}}
\newcommand{\linfor}{{\sf Linfor3D}}
\begin{document}
   \title{
Carbon-enhanced metal-poor stars: the most pristine objects?
\thanks{Based on observations obtained with the ESO Very Large
Telescope at Paranal Observatory, Chile (ID 087.D-0123(A).}
}

\author {
Spite M. \inst{1}\and
Caffau E.\inst{2,1}\and
Bonifacio P. \inst{1}\and
Spite F. \inst{1}\and 
Ludwig H.-G. \inst{2,1}\and
Plez B. \inst{3}\and 
Christlieb N. \inst{2} 
}

\institute {
GEPI Observatoire de Paris, CNRS, Universit\'e Paris Diderot, F-92195
Meudon Cedex France
\and
Zentrum f\"ur Astronomie der Universit\"at Heidelberg,
Landessternwarte, K\"onigstuhl 12, 69117 Heidelberg, Germany
\and
LUPM, CNRS, UMR 5299, Universit\'e de Montpellier II, F-34095 
Montpellier Cedex 05, France
}

%   \date{Received September 15, 1996; accepted March 16, 1997}

\authorrunning{Spite et al.}
\titlerunning{Carbon enhanced extremely metal-poor stars}

% \abstract{}{}{}{}{} 
% 5 {} token are mandatory
 
  \abstract
   {Carbon-enhanced metal poor stars (CEMP) form a significant proportion of the metal-poor stars, their origin is not well understood, and this carbon-enhancement appears in stars that exhibit different abundance patterns.}
   {Three very metal-poor C-rich turnoff stars were selected from the SDSS survey, observed with the ESO VLT (UVES) to precisely determine  the element abundances. In turnoff stars (unlike giants) the carbon abundance has not been affected by mixing with deep layers and is therefore easier to interpret.
     }
{
The analysis was performed with 1D (one dimensional)  LTE (local thermodynamical equilibrium) static model atmospheres.
When available, non-LTE corrections were applied to the classical LTE abundances. The 3D (three dimensional) effects  on the CH and CN molecular bands were computed
using  hydrodynamical simulations of the stellar atmosphere (\cobold) and are found to be very important.
}
   {
To facilitate a comparison with previous results, only 1D abundances are used in the discussion. The abundances (or upper limits) of the elements enable us to place these stars in different CEMP classes. The carbon abundances confirm the existence of a plateau  at  A(C)= 8.25 for $\rm [Fe/H] \ge -3.4$. The most metal-poor stars ($\rm[Fe/H] < -3.4$) have significantly lower carbon abundances, suggesting a lower plateau at $\rm A(C)\approx 6.5$.  Detailed analyses of a larger sample of very low metallicity carbon-rich stars are required to confirm (or refute) this possible second plateau and specify the behavior of the CEMP stars at very low metallicity.  
  }
{}


\maketitle

%
%---------------------------- Introduction -----------------------
\section{Introduction}

Carbon-enhanced extremely metal-poor stars (CEMP) have not yet been explained satisfactorily and obviously deserve more detailed investigations. At very low metallicity many stars are carbon enriched e.g.:  \citet{RossiBS99}, \citet{MarstellerBR05}, \citet{BeersC05}, \citet{FrebelCN06}, \citet{LucatelloBC06}. According to  Lucatello et al., more than 20\% of the metal-poor stars with $\rm [Fe/H] < -2.0$ exhibit $\rm [C/Fe] > +1.0$, and it has been shown  \citep{FrebelCN06} that the fraction of CEMP stars increases with the distance to the Galactic plane.

Moreover, the fraction of CEMP stars seems to increase when the metallicity decreases, and a significant carbon abundance has been considered an important factor to help the condensation of clouds into stars. Of the four stars known  with $\rm [Fe/H] < -4.5$, only one is known with a normal abundance pattern (no enhancement of C : SDSS\,J02915+172927  recently discovered  by \citet{CaffauBF11,CaffauBF12}), the other three are CEMP stars:
the CEMP fraction is very large, but the data-set (only four stars) is so limited that according to  \citet{YongNB13a} this is not statistically significant.

The carbon-enhanced stars exhibit various heavy-element abundance patterns: some are enriched  in heavy elements built by  both the ``s'' and the ``r'' processes (CEMP-rs stars), some are enriched  in only heavy elements built by the s-process (CEMP-s stars),  some others have a normal pattern of the heavy elements (CEMP-no stars).
\citet{MasseronJP10} have shown that generally CEMP-s and CEMP-rs stars have abundance patterns that suggest mass-transfer from a companion in the AGB (asymptotic giant branch).  A large percentage of the CEMP-s or -rs stars indeed shows radial velocity variations: this observed percentage is so large that it suggests \citep{LucatelloTB05} that all these stars are binaries. The AGB companions have different masses and abundances.

However, the abundance pattern of the CEMP-no stars is difficult to explain as caused by a mass transfer from an AGB companion.  For example, HE\,1327-2326 
\citep{FrebelAC05,FrebelCN06,AokiFC06} 
%(Frebel et al. 2005, 2006; Aoki et al. 2006)
is an extremely iron-poor star ([Fe/H]=--5.5) and has a ratio $\rm[Sr/Ba]>-0.4$. This ratio does not seem to be compatible with a production of Sr and Ba in an AGB companion. Moreover, so far only CS\,22957-027 exhibits direct evidence for duplicity \citep{PrestonS01}.  \citet{AokiFC06} have suggested that Sr might have originated from the r-process ($\rm[Sr/Ba]=-0.4$ is the limit compatible with the classical ``r'' process) or from a ``weak r process'' which is supposed to explain the high  [Sr/Ba] in some EMP stars with $\rm[Fe/H]<-3.0$. 
In addition, \citet{MasseronJP10} have found that the CEMP-no stars are more abundant at low metallicity and that the carbon enhancement in these stars is declining with metallicity, a trend not predicted by any of the current AGB models. 

Very few CEMP stars with $\rm[Fe/H]<-3.0$ have been studied, therefore observing  a few CEMP stars with $\rm [Fe/H] \le -3.0$ will help to specify the behavior of the CEMP stars at very low metallicity, and shed some light on the formation of the first stars.

%Table 1
\begin{table}
\begin{center}    
\caption[]{
Photometric data and preliminary estimates of [Fe/H] from SDSS spectra.
}
\label{tabstars}
\begin{tabular}{l@{~~}c@{~~}c@{~}c@{~}c@{~~}c@{~~}c@{}c}
%\begin{tabular}{lccccccc}
\hline
      &     &      &            &        & \Teff        &\Teff \\
star  & $g$ & $g-z$&$(g-z)_{0}$ & [Fe/H] &  H$_{\alpha}$&$(g-z)$\\       
      &     &      &            &  SDSS  &              &adopted\\       
\hline
%~~~\\
{SDSS } \\
J1114+1828  & 16.48 &0.414&  0.365& --3.5 & 6300 & 6200\\
J1143+2020  & 16.87 &0.385&  0.339& --3.5 & 6300 & 6240\\
J2209--0028 & 18.34 &0.457&  0.201& --4.5 & 6400 & 6440\\
\hline
\end{tabular}
\end{center}
\end{table}

%-------------------------- The star sample ------------------------
\section {Observations of the star sample} 
The ratios  [Fe/H] and [C/Fe] have been estimated for almost 30\,000 turnoff stars \citep{CaffauBFXh11}, from SDSS spectra \citep{Yorksdss00,Abazajian09}. The resolving power of these spectra is $R\approx 2000$. Three promising (C-rich) candidates with a value of [Fe/H] estimated to be below or equal to -3.5, have were selected (Table \ref{tabstars}). 
At present, the selection code \citep[see][]{CaffauBFXh11} selects only metal-poor turnoff stars (not giants); it is indeed more appropriate to analyze the C, N and O abundances in stars before they have undergone the first dredge-up which alters the C, N and O surface abundances by mixing with material processed in the star itself.
 
%In Fig. \ref{sp-sdss} the SDSS spectra are compared with synthetic spectra computed for scaled solar abundances. In all these stars the G-band is much stronger than the band expected with [C/Fe]=0.0. 

The spectra of the selected stars were then obtained at the VLT telescope with the UVES spectrograph \citep{DekkerDK00} in the course of two ESO periods. 
%Twenty hours in total were allocated for this program. 
We used a 1.5'' -wide slit and 2x2 on-chip binning. The resolving power (measured in the spectra) is $R \approx 39\,000$  with 2.8 pixels per resolution element. The S/N (signal-to-noise) ratio measured on the mean spectra (sum of the elementary spectra) is given in Table \ref{SN}. The spectra cover the ranges $\rm 376<\lambda <500\,nm $ (blue arm), $\rm 570<\lambda <750\,nm $, and  $\rm 767<\lambda <947\,nm $ (red arm).

%Table 2
\begin{table}
\begin{center}    
\caption[]{
g magnitude of the stars and signal-to-noise ratio of the mean spectrum at different wavelengths.
}
\label{SN}
\begin{tabular}{l@{~~}c@{~~}c@{~}c@{~}c@{~~}c@{~~}c@{}c}
%\begin{tabular}{lccccccc}
\hline
                    &  S/N    & S/N    &  S/N   &  S/N \\
star                &  ~~~395 nm & ~~~445 nm & ~~~588 nm & ~~~777 nm \\            
\hline
%~~~\\
{SDSS } \\
J1114+1828 &    50   &   70   &   60   &  100 \\
J1143+2020 &    40   &   60   &   60   &  100 \\
J2209--0028&    17   &   30   &   30   &   48 \\
\hline      
\end{tabular}
\end{center}
\end{table}

%Table 3
\section {Radial velocity measurements} \label{radvel}
\begin{table}
\begin{center}    
\caption[]{
Radial velocities in \kms ~measured on the individual UVES spectra. The date of the observation is given in the first column, the second column gives  the modified Julian date (MJD), the third column is the geocentric radial velocity ($\rm RV_{G}$) at the time of the observation, the fourth column lists the barycentric correction, and the two last columns provide the barycentric radial velocity $\rm RV_{B}$ and the corresponding error. An asterisk after the observation date means that the data correspond to the radial velocity measurement in the SDSS DR9 Science Archive Server.
}
\label{Vel}
\begin{tabular}{l@{~~~}c@{~~~}c@{~~~}c@{~~~}c@{~~}c@{~~}c@{}c@{}c }
date &   MJD&   $\rm RV_{G}$& Vcor& $\rm RV_{B}$& Err\\
\hline
J1114+1828\\
\hline
     4/05/2011&   55656.1673857&     234.9&     -15.5&     219.4&  1.   \\
    24/05/2011&   55705.9692687&     247.9&     -28.3&     219.6&  1.   \\
    25/05/2011&   55706.0053931&     247.6&     -28.4&     219.2&  1.   \\
    25/05/2011&   55706.0424672&     248.0&     -28.5&     219.5&  1.   \\
    03/06/2011&   55715.0406638&     247.9&     -28.9&     219.0&  1.   \\
    \\
   19/12/2007*&   54453.0024860&          &          &     231.3&  4.   \\                            
\\
J1143+2020\\
\hline
   22/06/2011&   55715.0808670&     260.1&     -28.2&     231.9&  1.   \\
   30/12/2011&   55925.3139552&     202.0&      27.4&     229.4&  1.   \\
   31/12/2011&   55926.2907973&     202.0&      27.2&     229.2&  1.   \\
   22/02/2012&   55979.2557883&     222.0&       7.8&     229.8&  1.   \\
   22/03/2012&   56008.1214818&     234.9&      -6.3&     228.6&  1.   \\
   \\
  20/03/2007*&   54179.3732200&          &          &     202.7&  4.   \\     
\\
J2209-0028\\
\hline
     3/06/2011&   55715.2467088&    -252.4&      29.0&    -223.4&  1.   \\
    28/07/2011&   55770.3553468&    -237.3&      13.7&    -223.6&  1.   \\
     3/08/2011&   55776.1416334&    -234.2&      11.7&    -222.5&  1.   \\
     3/08/2011&   55776.1780564&    -233.9&      11.6&    -222.3&  1.   \\
     3/08/2011&   55776.2150359&    -233.9&      11.5&    -222.4&  1.   \\
     3/08/2011&   55776.2520017&    -234.5&      11.3&    -223.2&  1.   \\
     3/08/2011&   55776.2895941&    -234.5&      11.2&    -223.3&  1.   \\
     3/08/2011&   55776.3270350&    -234.5&      11.1&    -223.4&  1.   \\
     5/08/2011&   55778.3313394&    -232.8&      10.2&    -222.6&  1.   \\
    25/08/2011&   55798.1409815&    -222.2&       1.1&    -221.1&  2.   \\
    \\
   20/09/2003*&   52902.2996680&          &         &     -222.2& 11.   \\
\hline
\end{tabular}
\end{center}
\end{table}

Since CEMP stars are suspected to often be binary stars, it is important to measure
 the radial velocity of the stars as precisely as possible at the time of the observation (Table \ref{Vel}). These radial velocities were measured on the blue spectra and the estimated error is about 1.0 \kms. 
The barycentric radial velocities are given in Table \ref{Vel}.
%do not show any significant variations during the time of the UVES observations (although the radial velocity of SDSS\,J1143+2020 seems to slightly decrease with time).
The radial velocities of SDSS\,J1114+1828 and SDSS\,J1143+2020 are significantly different from the values given in the SDSS DR9 science archive server (see Table \ref{Vel}), and therefore these two stars are very likely binaries.
In contrast, the radial velocity of SDSS\,J2209-0028 does not show any variation between 2003 and 2011.

%--------------- ANALYSIS -------------

\section {Analysis}
The effective temperature was derived from the $(g-z)_{0}$ color (Table \ref{tabstars}) using the calibration presented in Ludwig et al. (2008). The reddening correction is from \citet {SchlegelFD98}.  For comparison, we also determined effective temperatures by means of fitting synthetic line profiles to the wings of the observed H$\alpha$ lines. These temperatures agree well with those derived from photometry  (see Table \ref{tabstars}). \\ 
The gravity was fixed at $\log g = 4.0$, but we checked the ionization equilibrium of iron when some \Fed~ lines could be measured. The microturbulent velocity was assumed to be  1.3 \kms. In the three stars, the metallic lines are very weak and not sensitive to the adopted value.
We carried out a classical 1D LTE analysis using MARCS models 
\citep[e.g.][]{GustafssonBE75,GustafssonEE03,GustafssonEE08}.  The abundance analysis was performed using the LTE spectral line analysis code turbospectrum \citep{AlvarezP98,Plez12}.
The adopted solar abundances are given in Table \ref{abund}: Fe is taken from \citet{CaffauLS11}, the other elements are taken from \citet{LoddersPG09}.

The carbon abundance was determined by fitting the \Cu~line at 493.205 nm and the CH band at 422.4 nm (G band), and the nitrogen abundance by fitting the CN band at 388.8nm. The molecular data corresponding to the CH and CN bands are described in \citet{HillPC02} and \citet{Plez08}. 
The oxygen abundance was deduced from the IR triplet at 777\,nm.

In Table \ref{lines} all the lines used to derive abundances are listed with their wavelengths, excitation potentials, and oscillator strengths.   The word ``syn'' in place of the equivalent width means that the abundance was derived from spectral synthesis.\\

\subsection{1D computations - NLTE (non-LTE)  effects}
In extremely metal-poor stars such as the three stars we studied, NLTE effects are often significant: the collision rates are reduced (decrease of the electron density  with metal abundance), and since the radiation field is absorbed by a smaller number of metal atoms and ions, the photoionization rate tends to increase with decreasing metal abundance \citep{GehrenLS04}.

One \Cu~ line is visible in our spectra at 493.205nm. This high-excitation potential line is very sensitive to NLTE effects, and following \citet{BeharaBL10}, using the Kiel code \citep{steen84} and the carbon model atom described in \citet{SturH90}, the correction amounts to about --0.45dex in similar turnoff stars. This correction has been applied to the LTE value of the carbon abundance deduced from the \Cu~line.

\begin{figure}[ht]
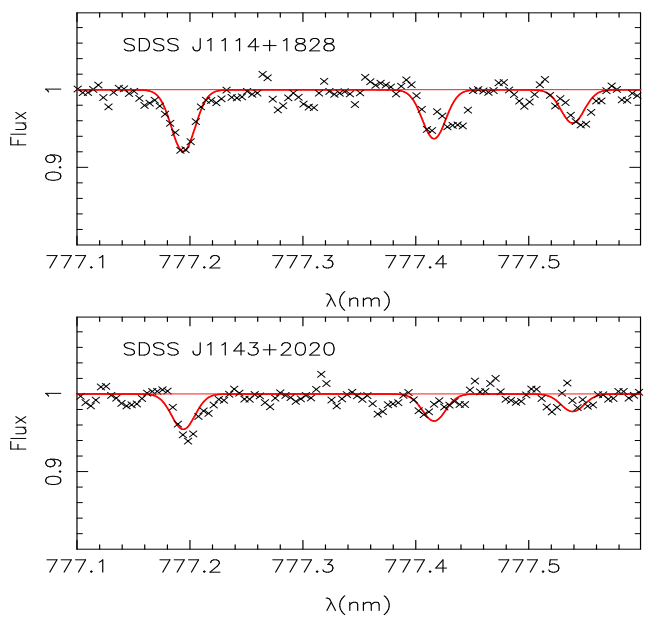

 \centering
\resizebox  {8.4cm}{4.0cm}
{\includegraphics {SDSSJ1114Oxy7768-7780.ps} }
\resizebox  {8.4cm}{4.0cm}
{\includegraphics {SDSSJ1143Oxy7768-7780.ps} }
   \caption[]{UVES spectrum in the region of the red \Ou~ triplet.} 
\label{oxy-a}
\end{figure}

The abundance of oxygen was computed from the red permitted \Ou ~triplet (Fig. \ref{oxy-a}). NLTE corrections to oxygen 
abundances were  computed using the Kiel code
\citep{steen84}, and
the model atom used in \cite{Paunzen99}.
We used the \citet{steen84} formalism to treat 
the collisions with neutral hydrogen.
We adopted  a scaling factor $\rm S_H=1/3$, as recommended
by \citet {caffau08}; the differences between the abundancescomputed 
with $\rm S_H=0$ and $\rm S_H=1$ is
less than 0.01 \citep[see][]{BeharaBL10}.

% Fig. 1

% Fig. 2
\begin{figure}[ht]
 \centering
\resizebox  {8.4cm}{4.2cm}
{\includegraphics {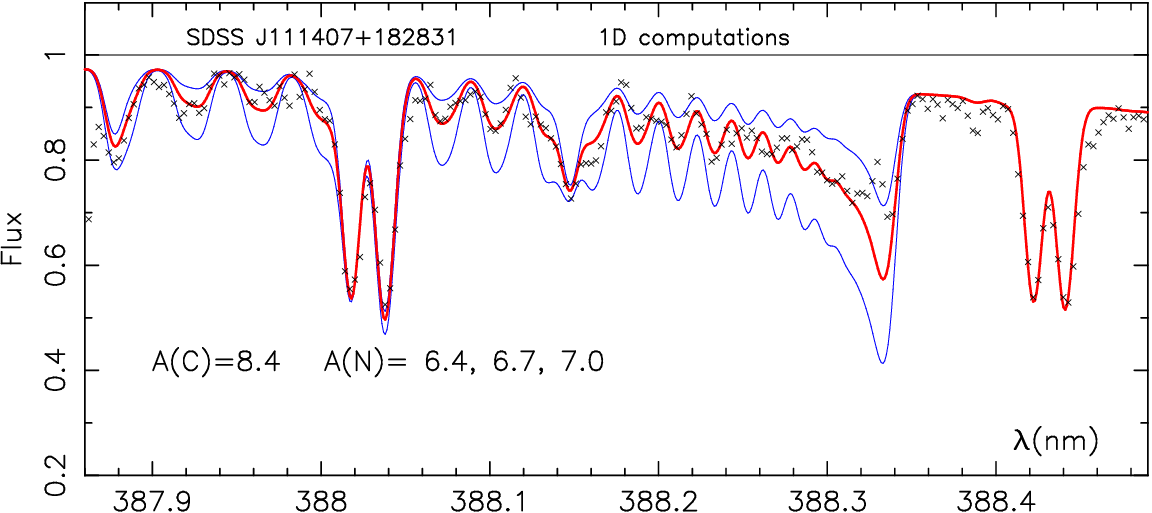} }
\resizebox  {8.4cm}{4.2cm}
{\includegraphics {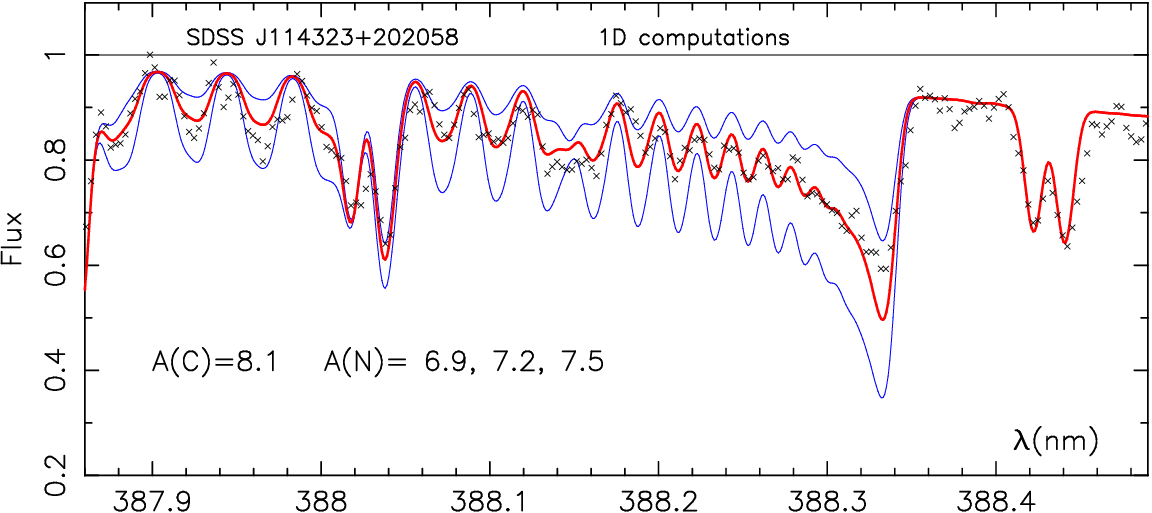} }
\caption[]{UVES spectrum in the region of the CN band compared to theoretical spectra.
The abscissa is the wavelength in nm, and the ordinate the relative flux.\\ 
For SDSSJ1114+1828 the theoretical spectra are computed with A(C{)} = 8.4 and A(N) = 6.4, 6.7 (best fit), 7.0,  and for SDSSJ1143+2020, with A(C{)} = 8.1 and A(N) = 6.9, 7.2 (best fit), 7.5.  }
\label{sp-cn}
\end{figure}

The abundances of the light metals from  sodium to calcium were corrected for NLTE effects following  \citet{AndrievskySK07, AndrievskySK08, SpiteAS12, Mashonkina-Mg13}.

The green magnesium triplet (multiplet 2) is beyond the wavelength range of our spectra and thus we deduced the magnesium abundance from multiplet 3 (at 383\,nm). 
\citet{Mashonkina-Mg13} has computed the NLTE correction for Mg taking into account inelastic collisions with neutral hydrogen atoms following \citet{BarklemBS12}.  From her computations on a small grid of models, the NLTE correction for multiplet 3 amounts to about +0.14\,dex for SDSSJ\,1114+1828, 0.10\,dex for SDSSJ\,1143+2020 and 0.25\,dex for SDSSJ\,2209-0028. 

The manganese abundances were derived by fitting synthetic spectra of the resonance lines (403\,nm) to observations, taking into account the hyperfine structure of the lines.  \citet{CayrelDS04}  underlined that the manganese abundance deduced from the resonance lines was 0.4\,dex lower (in giants) than the abundance deduced from the subordinate lines of \Mnu. Moreover  \citet{BonifacioSC09} have shown that there is a systematic difference between the ratio [Mn/Fe] in the turnoff stars and the giants even if only the resonance lines are taken into account. This behavior suggests a strong NLTE effect. The NLTE correction of the manganese abundance was computed for a grid of dwarf and subgiant models by \citet{BergemannG08}. Their most metal-poor model of turnoff star has a metallicity of -2.4 and for this metallicity the NLTE correction amounts to about +0.5 dex. This correction was applied to our LTE abundance.
 
\cite{LaiBJ08} and \cite{BonifacioSC09} have shown an offset between the chromium abundance derived from \Cru~ and \Crd~ lines in metal-poor stars, that points to NLTE effects.
\citet{BergemannC10} have computed the NLTE correction for a set of Cr lines in metal-poor stars. The atmospheric parameters of one star (G\,64-12) are close to those of our stars and thus we applied the NLTE correction found  for the stronger resonance line of \Cru ~in G\,64-12 (the only Cr line visible in our spectra): +0.44\,dex.
 
The abundances of the heavy elements Sr and Ba are generally sensitive to NLTE effects.  The abundance of Sr was corrected following \citet{AndrievskySK11}.
For Ba  it has been shown \citep{AndrievskySK09} that the NLTE correction depends almost only on [Ba/H] (the NLTE correction of the Ba lines is almost the same for two stars with different [Fe/H] but the same [Ba/H]).  Since  SDSS\,J1114+1828 and SDSS\,J1143+2020 are Ba-rich,  [Ba/H] in these two stars is between --1.3 and --1.8 dex. For these values of [Ba/H], the NLTE correction, following  \citet{MashonkinaGB99}, is close to zero (see their Table 4).

The line-by-line LTE abundances of the elements with the NLTE correction (when available) are given in Table \ref{lines}, and the resulting mean abundances are given in Table \ref{abund}.

%Table 4
\begin{table*}
\begin{center}    
\caption[]{
Elemental abundances in the observed stars from a {\bf 1D analysis}. Abundances of C(\Cu), O, Na, Mg, Al, Ca, Sr, and Ba have been corrected for NLTE effects (*), the LTE abundances for these elements are given line by line in Table \ref{lines}.
}
\label{abund}
%\begin{tabular}{l@{~~}c@{~~}c@{~}c@{~}c@{~~}c@{~~}c@{}c}
\begin{tabular}{|lr|rrrr|rrrr|rrrr|}
\hline
\multicolumn{2}{|c|}{-}&\multicolumn{4}{|c|}{SDSS\,J111407.07+182831.7}&\multicolumn{4}{|c|}{SDSS\,J114323.42+202058.0}&\multicolumn{4}{|c|}{SDSS\,J220924.74-002859.8}\\
\multicolumn{2}{|c|}{model}&\multicolumn{4}{|c|}{6200, 4.0, --3.3, 1.3}&\multicolumn{4}{|c|}{6240, 4.0, --3.3, 1.3}&\multicolumn{4}{|c|}{6440, 4.0, --4.0, 1.3}\\       
Species& $\epsilon_{\odot}$&$\epsilon$  &  [M/H]& [M/Fe]&N&$\epsilon$ &  [M/H]   & [M/Fe]&N&$\epsilon$ &  [M/H]   & [M/Fe]&N\\
\hline
C(CH) &   8.50&   8.40&   -0.10&   3.25&   & 8.10  & -0.40  &  2.75  &   &   7.10   &  -1.40&  2.56  &   \\
*C(\Cu)&  8.50&   7.95&   -0.55&   2.80& 1 & 7.75  & -0.75  &  2.40  & 1 &    -     &   -   &   -    &   \\
N(CN) &   7.86&   6.70&   -1.16&   2.19&   & 7.18  & -0.68  &  2.47  &   &    -     &   -   &   -    &   \\
 *O       &   8.76&   7.31&   -1.45&   1.90& 3 & 7.00  & -1.76  &  1.39  & 3 &   6.97   &  -1.79 &  2.17  & 3 \\
 *Na      &   6.30&   4.58&   -1.72&   1.63& 2 & 4.54  & -1.76  &  1.39  & 2 &$\le2.37$ &$\le-3.93$&$\le0.03$& 2 \\
 *Mg      &   7.54&   5.65&   -1.89&   1.46& 4 & 5.13  & -2.41  &  0.74  & 4 &   4.30   &  -3.24 &  0.72  & 3 \\
 *Al      &   6.47&   3.48&   -2.99&   0.36& 1 & 3.20  & -3.27  & -0.12  & 1 &    -     &    -   &   -    &   \\
 *Ca      &   6.33&   3.35&   -2.98&   0.37& 6 & 3.49  & -2.84  &  0.31  & 6 &   2.92   &  -3.41 &  0.55  & 5 \\
  Ti      &   4.90&   1.86&   -3.04&   0.31&11 & 2.03  & -2.87  &  0.28  & 11&    -     &    -   &   -    &   \\
 *Cr      &   5.64&   2.49&   -3.15&   0.20& 1 & 2.69  & -2.95  & -0.20  & 1 &    -     &    -   &   -    &   \\
 *Mn      &   5.37&   2.00&   -3.37&  -0.03& 1 & 2.25  & -3.12  &  0.03  & 1 &    -     &    -   &   -    &   \\
 \Feu     &   7.52&   4.17&{\bf-3.35}&   0.00&36 & 4.37  &{\bf-3.15}&  0.00  & 40&   3.56 &{\bf-3.96}&  0.0   & 6 \\
 \Fed     &   7.52&   4.18&   -3.34&   0.00& 3 & 4.24  & -3.28  & -0.13  & 3 &    -     &    -   &        &   \\
  Co      &   4.92&   2.40&   -2.52&   0.83& 3 & 2.26  & -2.66  &  0.49  & 3 &    -     &    -   &   -    &   \\
  Ni      &   6.23&   2.96&   -3.27&   0.07& 2 & 2.98  & -3.25  & -0.10  & 2 &    -     &    -   &   -    &   \\
 *Sr      &   2.92&   0.29&   -2.63&   0.72& 2 & 0.50  & -2.42  &  0.73  & 2 &$<-1.2$   &$<-4.12$&$<-0.16$& 1 \\
 *Ba      &   2.17&   0.44&   -1.73&   1.62& 3 & 0.86  & -1.31  &  1.84  & 3 &$<-0.74$  &$<-2.91$&$<+1.05$& 1 \\
  Eu      &  0.52&$<-1.10$&$<-1.62$&$<1.73$& 1 &$<-1.00$&$<-1.52$&$<1.63$& 1 &    -     &    -   &   -    &   \\
\hline    
$\rm[C/O]$  &     &       &  1.35  &       &   &       &  1.36  &        &   &          &  0.29  &        &   \\
$\rm[Ba/Eu]$&     &       & $> 0.1$&       &   &       &$> 0.4$ &        &   &          &        &        &   \\
$\rm^{12}C/^{13}C$&  &  & $> 60 $&       &   &       &$> 20 $ &        &   &          &        &        &   \\
\hline
\end{tabular}
\end{center}
\end{table*}

%Fig. 3
\begin{figure}!
 \centering
%\resizebox{7.2cm}{7.2cm}
\resizebox{\hsize}{!}
{\includegraphics{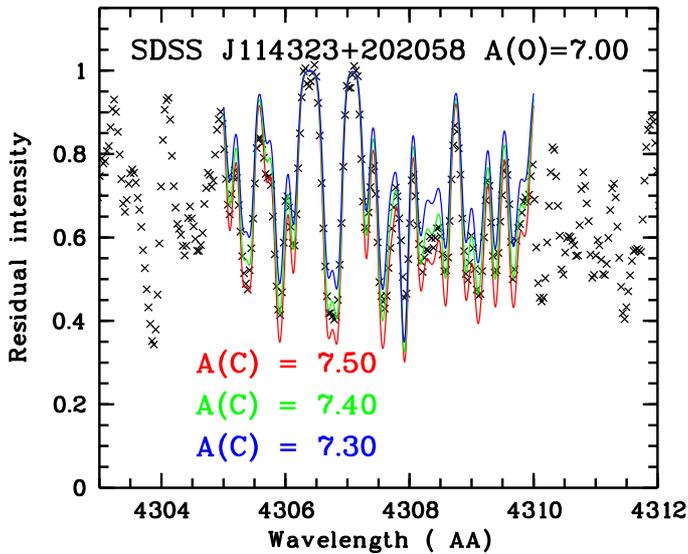} }
\caption[]{UVES spectrum of SDSS\,J114323+202058
in the core of the G band (black line), compared with two \linfor\ spectral
syntheses with the \cobold\ model, corresponding
to A(C)=7.30, 7.40 and 7.50.
The best fit is obtained with A(C)=7.40.} 
\label{gband1143}
\end{figure}

\subsection{Effects of three-dimensional hydrodynamical convection on the formation of molecular bands}

Classical models such as MARCS,
have a number of simplifying assumptions: the
atmosphere is approximated by a one-dimensional 
structure; all the physical quantities vary along
the vertical direction, but are constant in the
horizontal direction for any given depth; the atmosphere
is supposed to be in hydrostatic equilibrium; all
quantities are time independent; the stellar plasma
is supposed to be in LTE, i.e. the particle velocities
are given by a Maxwellian distribution at a local temperature
$T$, the atomic populations are given by Boltzmann factors, at
the {\em same} temperature $T$, ionization equilibria are also given
by Saha's law, at the {\em same} temperature.
In cool stars convection is a hydrodynamical, time-dependent
phenomenon, that needs to be treated in full three dimensional
geometry. 
To treat these effects it is necessary to use
hydrodynamical simulations, like those provided by
the \cobold code \citep{Freytag2012}, 
which are fully hydrodynamical, time dependent,  in a three
dimensional frame, yet retaining the LTE assumption. 
In this investigation we concentrated on the 
hydrodynamical effects on molecular bands, because they
have been shown to be extremely strong in metal-poor stars
\citep{ColletAT07,BeharaBL10,iJGH10}.

To estimate these effects we used a model
extracted from the latest version of the CIFIST
grid \citep{Ludwig-cifist09}.
The model has the parameters $\rm T_{eff}$/log g/[M/H]  
6300/4.0/--3.0, and consists of a box of $140 \times 140 \times 150$ 
points, corresponding to physical dimensions $26\times 26 \times 13$ Mm$^3$.
The opacity was derived from the MARCS suite 
\citep{GustafssonBE75,GustafssonEE03,GustafssonEE08} and 
was  binned into 12 opacity bins  following
the usual opacity binning scheme \citep{Nordlund83,LudwigPhD}.
\citet{BeharaBL10} noted a considerable difference between models computed with 6 opacity bins and those computed with 12 opacity bins.

The line formation in the three-dimensional structure
was computed using
\linfor \footnote{http://www.aip.de/$\sim$mst/Linfor3D/linfor\_3D\_manual.pdf}.
For the spectrum synthesis we used a selection of 20 snapshots,
chosen to be statistically independent and representative of the
overall characteristics of the hydrodynamical simulation.   
For comparison, \linfor\ also integrates the transfer equation through two reference one-dimensional structures: \\
-- an \xx\ model \citep[see][]{CaffauL07} that uses the same 
microphysics as \cobold ; and \\
-- a \mD~ model, that is obtained by
averaging the three dimensional structure over surfaces of constant Rosseland optical depth and over time. 

Following \citet{CaffauL07},  we defined the
3D  correction  as the difference in the abundance 
derived from the full 3D synthesis and that derived
from the \xx\ model (3D -- \xx ).
It is also useful to consider the difference between
the abundance derived from the 3D model and the \mD\ model.
Since the two models have by construction the same
mean temperature distribution, this difference helps
to single out the effect of temperature fluctuations. 

Computing molecular bands is CPU-intensive, 
since many lines need to be treated. The G-band is very challenging
since it extends over almost 20\,nm.
The  data for the G-band (CH AX electronic system)
were taken from \citet{PlezCo05} and \citet{Plez08}.
%We used three different methods to estimate the 3D correction
%of the G-band;
%\begin{enumerate}
%\item as in \citet{BeharaBL10} we selected 
%4 isolated lines on the blue wing of the band, for which
%an equivalent width could be measured, and computed 
%the 3D corrections or these 4 lines;
%\item as in \citet{CaffauBF12} 
%we computed 4 artificial lines of  excitation
%energies from 0 to 1.4 eV  and 
%f-value to sample strong and weak lines
We computed the full 3D synthesis of the G band core
for various carbon abundances and compared them 
with the observed spectrum. The result for
SDSS\,J114323+202058 is shown in Fig.\ref{gband1143}.
The best fit is obtained for A(C)=7.4 and as a consequence, the 1D to 3D correction is about -0.7\,dex.
%of the core of the
%G band for various carbon abundances and compared them to
%the observed spectrum. The result for 
%SDSS\,J114323+202058 is shown in Fig.\ref{gband1143}.

%
%The three methods provide more or less the same result and
%{\bf we conclude that the 3D correction for the carbon
%abundance is about --0.7 dex.} 

\begin{figure}[ht]
 \centering
\resizebox  {8.4cm}{4.2cm}
{\includegraphics {cn1143p2020.ps} }
\resizebox  {8.4cm}{4.2cm}
{\includegraphics {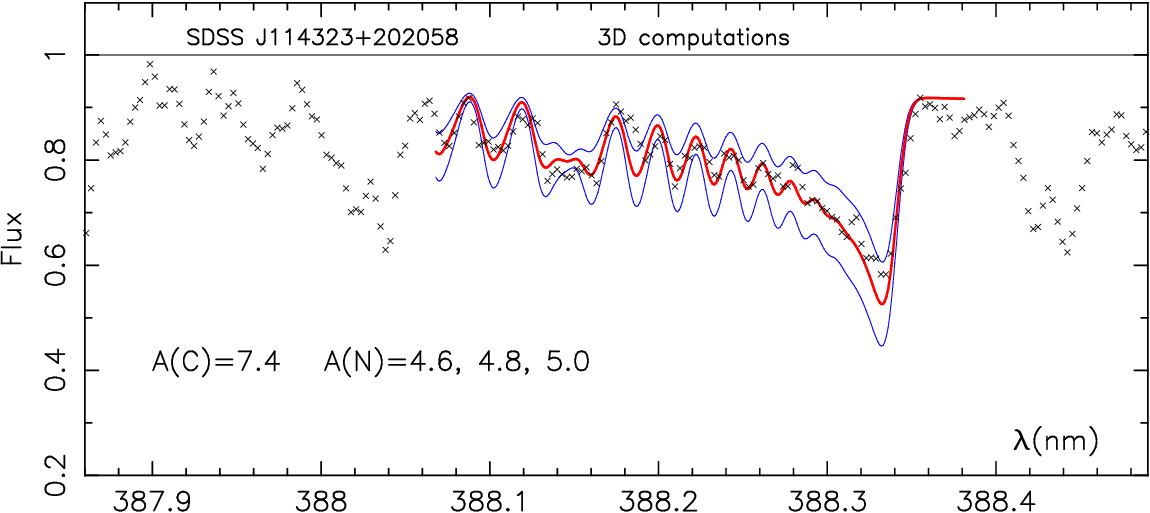} }
   \caption[]{UVES spectrum in the region of the CN band 
of SDSS\,J1143+2020. The abscissa is the wavelength in nm, and the ordinate
is the relative flux.\\
In the top panel the observed spectrum (crosses) is compared to 1D theoretical spectra  computed with  A(C{)} = 8.1 (best fit for the CH band in 1D) and A(N) = 6.9, 7.2 (best fit), 7.5.\\ 
In the bottom panel the observed spectrum is compared to 3D theoretical spectra
computed with A(C) = 7.4 (best fit for the CH band in 3D, see Fig. \ref{gband1143})  and A(N)= 4.6, 4.8 (best fit) and 5.0.}
\label{3Dprofil}
\end{figure}

For the nitrogen abundance we used the CN  
BX [\,B$^2\Sigma^+$ --X$^2\Sigma^+$ (0--0)\,] violet band at
388\,nm.
We used the molecular data from \citet{HillPC02} and 
\citet{Plez08}, but considering the limited spectral
range covered by the band (about 0.5\,nm), it is
feasible to synthetise the whole band, as shown in Fig.\,\ref{3Dprofil}.
The 3D effect seems to be much more important 
for this band than for the CH band.
This band is formed very close to the surface of the star 
and the nitrogen abundance deduced from this band is also 
strongly affected by the quantity of oxygen atoms bound at this level 
in the CO molecule. As a consequence, we found for these oxygen-rich 
stars, a huge correction: about --2.4 dex. 
This correction (if confirmed) would almost annihilate the 
nitrogen enhancement in the CEMP turnoff stars. 
These 3D corrections affect not only the three turnoff 
stars we studied but all CEMP stars in the literature 
where these 3D effects have generally been neglected.
These effects could be so significant that they could lead to a 
revision of the interpretation of the abundances in 
these stars (type of AGB stars responsible for the abundance anomalies).
In Figure \ref{3Dprofil} we compare the 1D and the 3D profiles of the CN band in SDSS\,J1143+2020 with different abundances.

% Fig. 4

To confirm this very large correction of the nitrogen abundance, 
it would be very important to compare the abundance 
deduced from the CN band and from the NH band, which is 
practically independent of the oxygen abundance.
But the NH band is not in the observed spectral range of the 
very faint stars studied in the present paper.
We will study this effect, which is very sensitive to the most external layers of the models,
 in another sample of CEMP stars with known CN and NH bands in a future investigation.

Since the aim of the present paper is to enlarge the sample 
of the abundance patterns in CEMP stars, 
we  discuss  in the remainder of this paper the 1D abundances 
to facilitate a comparison with the previous analyses, but
with the caveat that the abundance of C and N could be strongly 
reduced.  

\subsection{SDSS\,J1114 +1828  and SDSS\,J1143+2020} 

Both stars are very metal-poor ($\rm[Fe/H]<-3.0$) and exhibit a very strong enhancement of C, N, O and Mg (Table \ref{abund}). This enhancement tends to decrease with the atomic number. The CN feature is displayed in  Fig. \ref{sp-cn}.

The carbon abundance deduced from the \Cu ~line is systematically lower by about 0.4 dex than the abundance deduced from the CH band. This difference has also been observed by  \citet {BeharaBL10}. A 3D computation of the  CH band and the \Cu ~lines does not remove this discrepancy, but in this case the abundance of C derived from the \Cu ~line is higher than the abundance derived from the CH band.
Since in the literature the carbon abundance is generally deduced from 1D computation of the CH band  we used the C (CH) abundance in the forthcoming discussion.

The overabundance of C, N, and O in these stars is generally attributed to a production in an AGB companion that transfers some processed material to the observed star. 
This hypothesis is reinforced by the fact that the radial velocity of both stars have been found to be variable in section \ref{radvel}.
%Several models have been proposed in particular by Hirschi et al. \cite{HFL06}, and Meynet at al. \cite{MEM06}.
In Figure \ref{cfenfe} we plot these two new CEMP turnoff stars in a diagram [N/Fe] vs. [C/Fe] for comparison with the values found for other CEMP dwarfs and turnoff stars 
\citep{SivaraniBB06,BeharaBL10,MasseronJL12}
%(Sivarani et al., \cite{SBB06}, Behara et al., \cite{BBL10}, Masseron et al., \cite{MJL12}) 
and the predictions of \cite{Herwig04} computed for AGB stars with $6 M_{\odot}$ HBB (Hot Bottom Burning model) and $2 M_{\odot}$ non-HBB. Like most of the CEMP turnoff stars, SDSS\,J111407+182831 and SDSS\,J1143+2020 lie in the region delimited by these two models \citep[see][]{SivaraniBB06}.

However, we must have in mind that in a 3D analysis all stars in this graph would have to be shifted toward lower values of [C/Fe] (--0.5 dex) and mainly much lower values of [N/Fe]. This will be discussed in a follow-up paper.

% Fig. 5
\begin{figure}[ht]
 \centering
\resizebox  {7.0cm}{7.0cm}
{\includegraphics {cfe-nfe.ps} }
\caption[]{[N/Fe] vs. [C/Fe] for CEMP dwarfs and turnoff stars from the recent literature  
\citep[~diamonds]{SivaraniBB06}, 
\citep[~stars]{BeharaBL10}, 
\citep[~open circles]{MasseronJL12}, 
\citep[~asterisk]{ChristliebBB02},  and for our stars SDSS\,J111407+182831 and SDSS\,J1143+2020 (filled circles). The AGB models predicted by \citet{Herwig04}, for [Fe/H]=--2.3 and initial masses $6 M_{\odot}$ (HBB) and $2 M_{\odot}$ (non-HBB) are labeled by triangles.
} 
\label{cfenfe}
\end{figure}

% Fig. 6
\begin{figure}[ht]
 \centering
\resizebox  {7.0cm}{7.0cm}
{\includegraphics {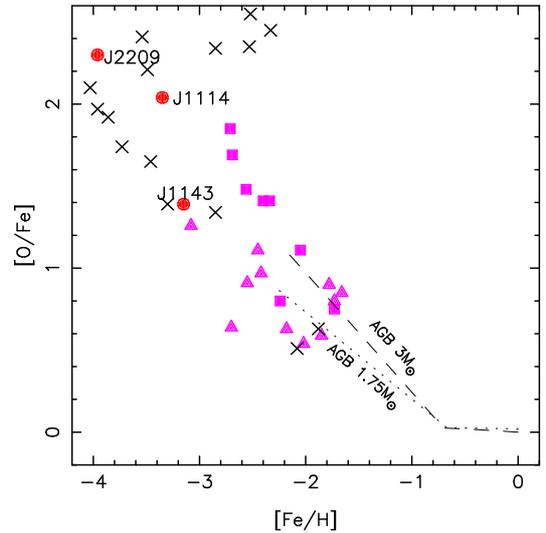} }
   \caption[]{[O/Fe] vs. [[Fe/H]] for CEMP-s  stars (triangles), CEMP-rs stars (squares), CEMP-no stars (crosses) following \citet{MasseronJP10}. Our stars are represented by red full circles. The predictions of \citet{KarakasL07} for pollution by $1.75 M_{\odot}$ and $3 M_{\odot}$   AGB are indicated as dotted and dashed lines. } 
\label{ofe}
\end{figure}

% Fig. 7
\begin{figure}[ht]
 \centering
\resizebox  {8.4cm}{7.0cm}
{\includegraphics {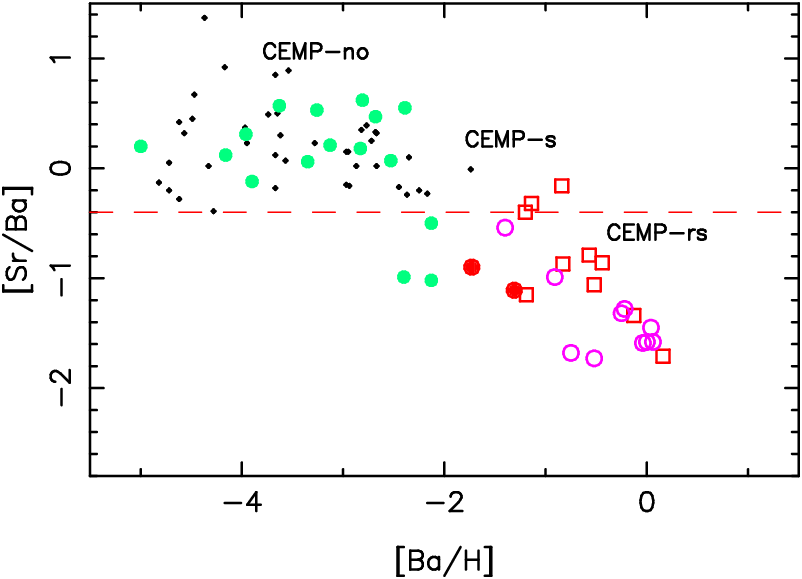} }
\caption[]{[Sr/Ba] vs. [Ba/H] for normal metal-poor stars (black dots) following \citep{AndrievskySK11}, and CEMP stars following \citet{Frebel10} and \citet{CescuttiChiap13}. 
Green filled circles represent CMP-no stars, open red squares CMP-s stars, and open magenta circles  
CMP-rs stars. The red dots represent SDSS\,J111407+182831 and SDSS\,J1143+2020. 
In case of mass transfer from an AGB companion $\rm[Sr/Ba] < -0.4$ (dashed line). Our stars fulfill this requirement.} 
\label{SrBa}
\end{figure}

\subsubsection{Heavy elements Sr, Ba, Pb, and class determination of the stars}

  With a Ba overabundance $\rm[Ba/Fe]>1.0$, the two stars SDSS\,J111407+182831 and SDSS\,J1143+2020 belong  \citep{BeersC05}, to the class of the CEMP-s stars or to the CEMP-rs stars, depending on the abundance of Eu ($\rm[Ba/Eu]>0.5$ in CEMP-s stars and $\rm 0.0<[Ba/Eu]<0.5$ in CEMP-rs stars). The europium line is not visible in these two CEMP turnoff stars, and the low S/N of the spectra in the region of the main Europium line (412.9~nm) does not allow a very restrictive estimate of the Eu abundance. 

\citet{MasseronJP10} have proposed a classification based on [Ba/Fe] alone, when [Eu/Fe] is not available. From their Figure 1, the stars are all CEMP-rs when $\rm [Ba/Fe]>2.1$. But when $\rm 1.6<[Ba/Fe]<1.9$ like in our stars,  there is about the same number of CEMP-s and CEMP-rs stars. 

On the other hand, \citet{MasseronJP10} have shown that, for the same metallicity, the CEMP-rs stars have a higher value of [O/Fe] than the CEMP-s stars. In Fig. \ref{ofe} we have plotted [O/Fe] vs. [Fe/H] for the stars of \citet{MasseronJP10}, with $\rm [Fe/H]< -1.5$ (their Fig. 18) and for our stars; the dashed and dotted lines in this figure represent the predictions of  \citet{KarakasL07} for AGB of $1.75 M_{\odot}$ and  $3 M_{\odot}$. This graph supports the idea that  CEMP-rs stars have been polluted by stars with a higher mass AGB than CEMP-s stars \citep{MasseronJP10} and that  SDSS\,J111407+182831 with its very high oxygen abundance could be classified as a  CEMP-rs, and SDSS\,J1143+2020 as a CEMP-s star.

In Fig. \ref{SrBa} we have plotted [Sr/Ba] as a function of [Ba/H] for different types of CEMP stars. The abundances of Sr and Ba have been taken from \citet{Frebel10}. All CMP-s and CMP-rs stars have a known Europium abundance. We adopted the definitions of \citet{SivaraniBB06}: we called CEMP-no stars all CEMP stars with $\rm [Ba/Fe]<1.0$ (CEMP-no or CEMP-low-s), the CEMP-s and CEMP-rs stars have $\rm [Ba/Fe]>1.0$ with $\rm [Ba/Eu]>0.5$ for the CEMP-s and $\rm 0<[Ba/Eu]<0.5$ for the CEMP-rs stars. The small dots represent the ``normal'' EMP stars \citep{AndrievskySK11}.      
The ratio [Sr/Ba] in the CEMP-s and CEMP-rs stars is much lower than in the classical EMP stars. This low ratio is generally compatible with a mass transfer from an AGB companion $\rm([Sr/Ba]<-0.4$). 
In this figure, both SDSS\,J111407+182831 and SDSS\,J1143+2020 are located in the region corresponding to both the CEMP-s and CEMP-rs stars, their abundances of Sr and Ba are compatible with a mass transfer from an AGB (unlike the majority of the CEMP-no stars).

%{\bf  CEMP-s, CEMP-rs, and CEMP-no stars are located in different regions of a diagram  [Sr/Ba] vs. [Ba/H]. In Fig. \ref{SrBa},  SDSS\,J1114+1828 and  
%SDSS\,J1143+2020 are rather in the region of the CEMP-s stars.
%Their abundances of Sr and Ba  are $<-0.4$ are not incompatible with a mass transfer from an AGB (unlike the majority of the CEMP-no stars). 
%}
%But a determination of the europium abundance in these stars would be necessary to decide between the class s and the class rs.

Following \citet{MasseronJP10}, when $\rm [Fe/H] < -3.0$, the CEMP-no stars ($\rm [Ba/Fe]<+1.0$) are much more numerous  than the CEMP-s or CEMP-rs stars. These authors note that neither CEMP-rs  nor CEMP-s stars have been discovered below $\rm [Fe/H] < -3.2$. Thus SDSS\,J111407+182831 with [Fe/H]=--3.35 is to date the most metal-poor CEMP-s or CEMP-rs star. 

\begin{figure}[ht]
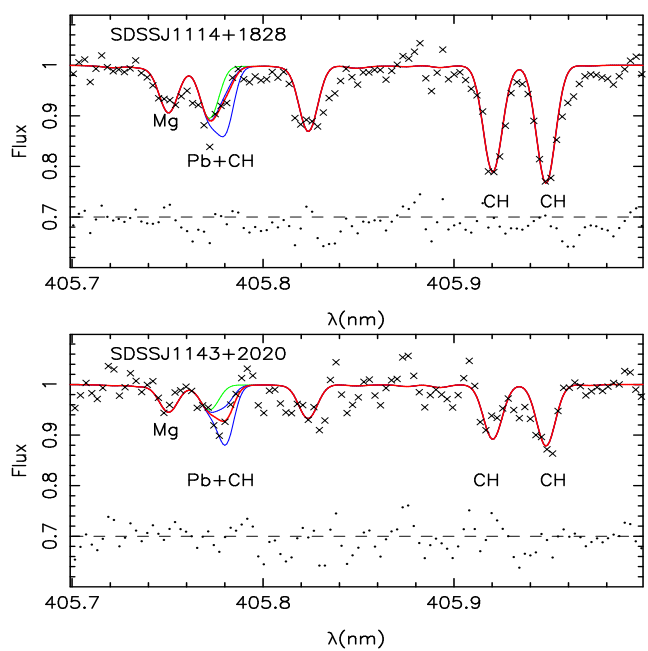

 \centering
\resizebox  {8.4cm}{4.2cm}
{\includegraphics {SDSSJ1114-Pb.ps} }
\resizebox  {8.4cm}{4.2cm}
{\includegraphics {SDSSJ1143-Pb.ps} }
   \caption[]{Spectra of SDSS\,J1114+1828 and SDSS\,J1143+2020 in the region of the Pb line. The synthetic spectra have been computed without Pb (green line) and with A(Pb)= 1.5 and 2.1 (thin blue lines). The best fits (red lines) are obtained with A(Pb)=1.6 for SDSS\,J1114+1828 and A(Pb)=1.8 for SDSS\,J1143+2020.} 
\label{pb}
\end{figure}

Several carbon-rich stars CEMP-s and CEMP-rs stars are also lead-rich (see e.g.: \citet{CohenCQ03}, 
\citet{BarbuySS05}, \citet{BisterzoGS06}). We tried to detect Pb in SDSS\,J1114 +1828  and SDSS\,J1143+2020 (Fig. \ref{pb}). The best fits were obtained with  $\rm A(Pb)=1.6\pm 0.3$  for SDSS\,J1114+1828, and $\rm A(Pb)=1.8 \pm 0.2$ for SDSS\,J1143+2020. The corresponding equivalent widths are respectively about 4 and 6 m\AA. In this region of the spectrum the S/N is around 50 (see Table \ref{SN}), therefore from the \citet{Cayrel88} formula, the error on the equivalent width is lower than 2m\AA. For SDSS\,J1143+2020 the detection of Pb is highly significant ($>95\%$).
In both stars $\rm[Pb/Fe] \approx +3.0$, a value very close to the values of [Pb/Fe] generally found from LTE computations in similar stars  \citep{BisterzoGS06}.
But the Pb abundance has to be corrected for NLTE effects which, following \citet{MashonkinaRF12}, are strong in metal-poor stars. These authors have not computed the correction for turnoff stars with a metallicity as low as those of SDSS\,J1114+1828 and SDSS\,J1143+2020,  but since the correction increases with the temperature and with decreasing metallicity, it is at least equal to +0.3 according to their Table 1,

%% Fig. 8
%\begin{figure}[ht]
% \centering
%\resizebox  {7.1cm}{5.9cm}
%{\includegraphics {srba.ps} }
%   \caption[]{[Sr/Ba] vs. [[Ba/H]] for normal metal-poor stars (dots), CEMP-s  stars or CEMP-rs stars (red squares), CEMP-no stars (crosses). SDSS\,J111407+182831 and SDSS\,J1143+2020 are represented by red full circles. In case of mass transfer from an AGB companion $\rm[Sr/Ba] < -0.4$ (dashed line). Our stars fulfill this requirement as all the CEMP-s and CEMP-rs stars.} 
%\label{srba}
%\end{figure}

% Fig. 8

\subsubsection{$\rm^{12}C/^{13}C$ ratio}
\begin{figure}[ht]
 \centering
\resizebox  {8.4cm}{4.2cm}
{\includegraphics {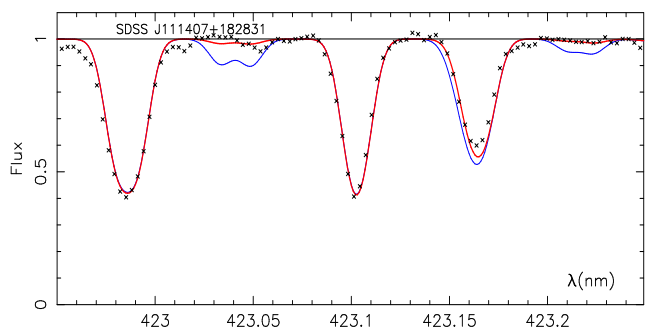} }
   \caption[]{UVES spectrum in the region of the CH band in SDSS\,J111407+182831 compared to theoretical spectra computed with $\rm A(^{12}C{)}$=8.4 and $\rm^{12}C/^{13}C$ equal to either 10 (thin blue line) or 60 (thick red line). The abscissa is the wavelength in nm, and the ordinate the relative flux. The $\rm^{13}C$ feature is not visible in the observed spectrum, and we derived for this star $\rm^{12}C/^{13}C \ge 60$.} 
\label{sp-c13}
\end{figure}

The  $\rm^{13}C$ features are not visible in the observed spectra (Fig.\ref{sp-c13}). We deduced lower limits: $\rm^{12}C/^{13}C \ge 60$ for SDSS\,J111407+182831 and $\rm^{12}C/^{13}C \ge 22$ for SDSS\,J1143+2020. An example of the  $\rm^{13}C$ feature is displayed in Fig. \ref{sp-c13}.
These lower limits are compatible with the values found in the CEMP-s and CEMP-rs stars \citep[see][ their Fig.17]{MasseronJP10}. \citet{MasseronJP10} have remarked that, although the mechanism responsible for the N production may be attributed to CBB (cool bottom burning) in the framework of AGB evolution, no current AGB models reproduce the observed trend of $\rm \log ^{12}C/^{13}C $ vs. [C/N].

% Fig. 9

% Fig. 10
\begin{figure}[ht]
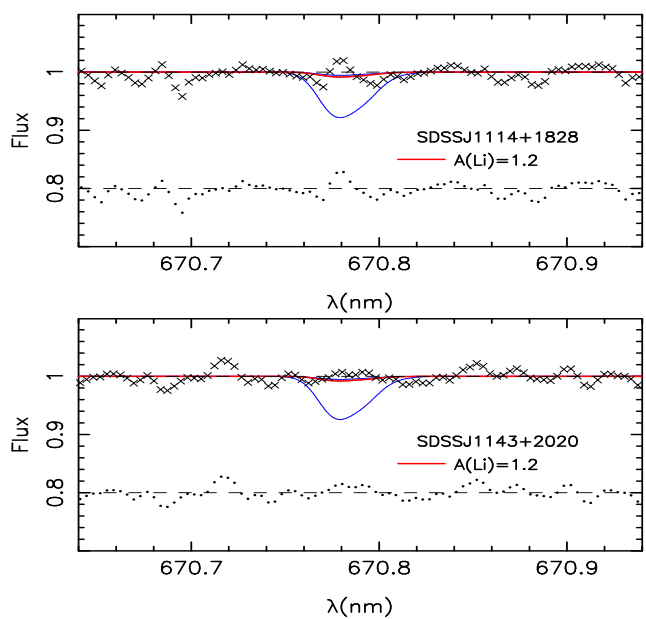

 \centering
\resizebox  {8.4cm}{4.0cm}
{\includegraphics {SDSSJ1114-lithium.ps} }
\resizebox  {8.4cm}{4.0cm}
{\includegraphics {SDSSJ1143-lithium.ps} }
   \caption[]{UVES spectrum in the region of the lithium feature. The thin blue line corresponds to the lithium feature computed with the classical abundance A(Li)= 2.2,  and the thick red line to the best fit (A(Li)=1.2 in both cases). The dots at the bottom of the figures represent the difference between this best fit and the observed spectrum (shifted upward by 0.8).
 We adopted in both cases $\rm A(Li) < 1.5$.
 } 
\label{Li-a}
\end{figure}

\subsubsection{Lithium abundance}
In normal (non C-enhanced) turnoff stars with a metallicity higher than $\rm[Fe/H]=-3.0$, the lithium abundance A(Li) is generally constant, independent of the temperature and metallicity and equal to about 2.2 dex defining a plateau \citep[see e.g.][]{SpiteSB12}. But below $\rm[Fe/H]=-3.0$ there is a meltdown of the plateau 
\citep[e.g.][]{BonifacioMS07,GonzalezBL08,SbordoneBC10} and a large part of the normal very deficient stars have a lithium abundance lower than 2.2.

In the CEMP turnoff stars, even with a metallicity higher than $\rm[Fe/H]=-3.0$, the lithium abundance is often lower than the value of the plateau. 
In the spectra of SDSS\,J1114+1828 and SDSS\,J1143+2020 the lithium line is not visible (Fig. \ref{Li-a}). We derived in both cases $\rm A(Li)\leq 1.5$, a lithium abundance well below the value of the plateau. 

The low abundance of Li in CEMP (s and rs)   stars is discussed by \citet{MasseronJL12}. Briefly, before the transfer of material from the AGB companion, the lithium abundance in the observed star should be the same as in normal metal-poor turnoff stars, but as more and more AGB material is transferred to the atmosphere of the star, the carbon abundance  increases while the lithium abundance decreases because in most cases the lithium abundance in the AGB star is lower than the lithium plateau.

% Fig. 11
\begin{figure}[ht]
 \centering
\resizebox  {8.4cm}{4.0cm}
{\includegraphics {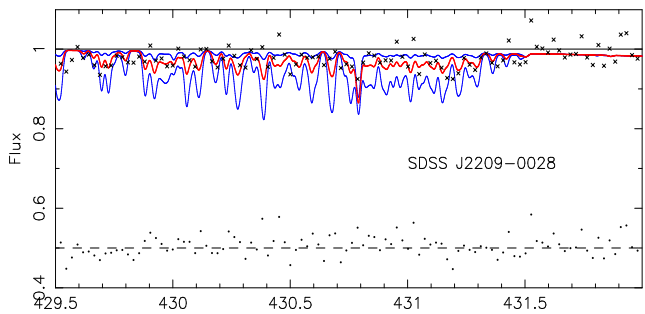} }
   \caption[]{UVES spectrum in the region of the CH band in SDSS\,J2209-0028. The thin blue lines correspond to the computation of the CH band with an abundance A(C)=6.5 and 7.5. The thick red line corresponds to A(C)=7.1 (best fit) and the dots at the bottom of the figure represent the difference between the synthetic spectrum computed with A(C)=7.1  and the observed spectrum (shifted upward by 0.5).} 
\label{CH2209}
\end{figure}

% Fig. 12
\begin{figure}[ht]
 \centering
\resizebox  {8.4cm}{4.0cm}
{\includegraphics {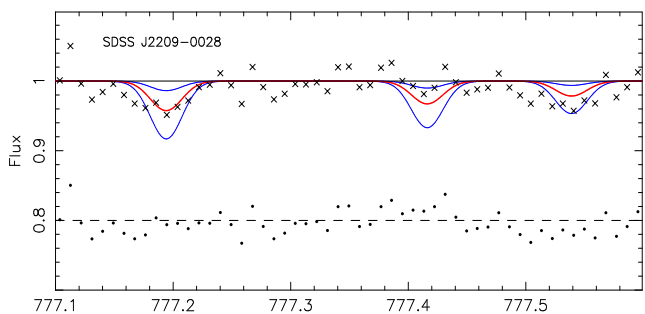} }
   \caption[]{Thin blue lines correspond to synthetic spectra computed with A(O)= 6.5 and 7.5. and the thick red line (best fit) has been obtained with A(O)=7.1. The crosses represent the observed spectrum (binned x2) and the dots at the bottom of the figure are the difference between the best fit and the observed spectrum (shifted upward  by 0.8).} 
\label{oxy-b}
\end{figure}

\subsection{SDSS\,J220924-002859}
SDSS\,J220924-002859 with $\rm [Fe/H]\approx -4$ is the most metal-poor star of our sample. Compared to the other two stars its carbon abundance is also ten times lower. But it is also the faintest star of our sample (Table \ref{tabstars}) and the S/N of the spectra is much lower (Table \ref{SN}). As a consequence the abundance of only few elements could be measured in this star.

%\subsubsection{Carbon abundance in SDSS\,J220924-002859 and in the dwarf CEMP stars}
The 1D carbon abundance in SDSS\,J220924-002859 was estimated to be $\rm A(C)=7.0\pm 0.2dex$  (Fig \ref{CH2209}), with the 3D correction we derived $\rm A(C)=6.5\pm 0.2dex$. However, we remark that the S/N in this star is so poor that the CH band in the spectrum is at the limit of the detection and it would be very interesting to obtain a new better spectrum in this region.

%\subsubsection{Oxygen abundance in SDSS\,J220924-002859 and the class of the star}
With [O/Fe]=2.10, SDSS\,J220924-002859 is the most oxygen-rich star in our sample (Fig. \ref{oxy-b}). 
The barium line is not visible in our spectrum and we derived, within the LTE hypothesis, an upper limit $\rm[Ba/Fe]<0.75$  (this upper limit becomes $\rm[Ba/Fe]<1.05$ when the NLTE effects are taken into account). It should thus belong to the CEMP-no stars (defined by $\rm[Ba/Fe]_{LTE}<+1$ see also the Fig. \ref{SrBa}).

The star is very oxygen-rich, and in Fig. \ref{ofe} it indeed appears in the region of the 
CEMP-no stars.
SDSS\,J220924-002859 belongs to this class, as do most of the extremely iron-poor carbon-rich stars \citep{MasseronJP10}. 

% Fig. 13
\begin{figure}[ht]
 \centering
\resizebox  {8.4cm}{4.0cm}
{\includegraphics {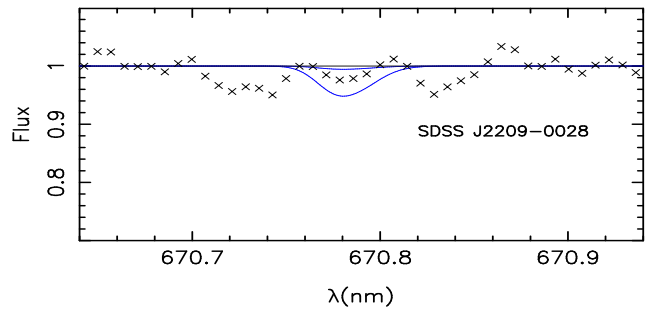} }
\caption[]{UVES spectrum in the region of the lithium feature. The thin blue lines correspond to  the lithium feature computed with the abundance A(Li)=1.2 and 2.2. The noise is important in this region of the spectrum and therefore we derive $\rm A(Li)\le 2.2$.
} 
\label{Li-b}
\end{figure}

%\subsubsection{Lithium abundance in SDSS\,J220924-002859}
In SDSS\,J220924-002859 the best fit is obtained with $\rm A(Li)=1.8\pm 0.4$ (Fig. \ref{Li-b}). 
%This value seems again lower than the value of the plateau. 
The noise is very high in this region of the spectrum and the abundance of the plateau
A(Li)=2.2 cannot be excluded.

\section {Discussion and Conclusion}

% Fig. 14
\begin{figure}[ht]
 \centering
%\resizebox  {8.4cm}{4.0cm}
\resizebox  {9.0cm}{5.0cm}
{\includegraphics {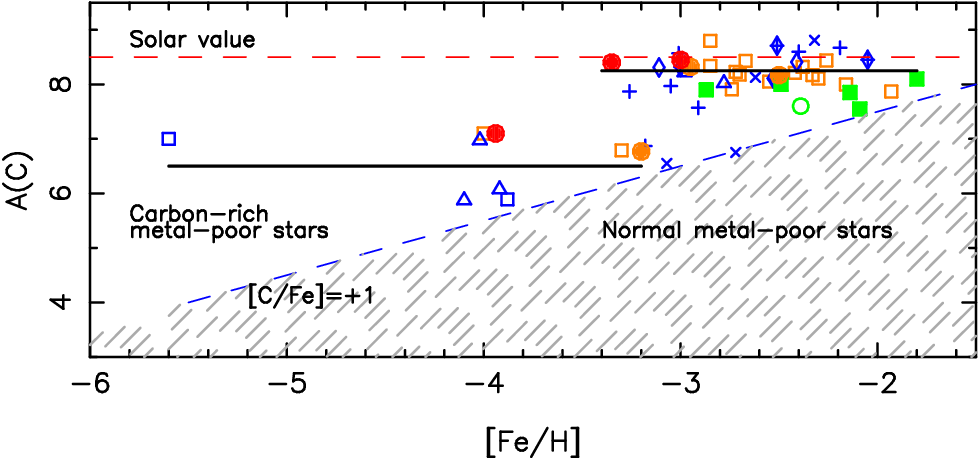} }
   \caption[]{Abundance of carbon A(C) vs. [Fe/H] in dwarfs and turnoff CEMP stars,  
following 
\citet[their Table 4]{SivaraniBB06}, [orange open squares]), \citet{FrebelAC05,FrebelJB07} [blue open squares], \citet{ThompsonIB08} [green open circle], 
\citet{AokiBS08} [blue open diamonds], 
\citet{BeharaBL10} [full orange circles], 
\citet{PlaccoKB11} [blue $\times$), 
\citet{CarolloBB12} [blue $+$],
\citet{MasseronJP10,MasseronJL12} [full green squares], 
\citet{YongNB13b}  [blue open triangles]. 
Our measurements are represented with full red circles.
The dashed blue line (representing $\rm[C/Fe]=+1$) separates the region of the carbon-rich metal-poor stars from the region of the normal metal-poor stars (hatched).  When [Fe/H] is higher than --3 the carbon abundance is almost constant and close to 8.25 \citep[see also][]{MasseronJP10}. For 
$\rm [Fe/H] < -3.4$ the carbon abundance drops suddenly to about 6.5. However, it is difficult to decide whether at very low metallicity A(C) is also constant, or whether the line A(C)=6.5 represents an upper limit of the carbon abundance in these extremely metal-poor star \citep{MasseronJP10}.  
} 
\label{cdwarfs}
\end{figure}

% DIS   CONCLUSION

%$\bullet$ 
In Fig.\ref{cdwarfs} we plotted  the recent measurements of the abundance of C in CEMP dwarf and turnoff stars as a function of [Fe/H], for $\rm [Fe/H]<-1.8$ collecting the data of  \citet[their Table 4]{SivaraniBB06}, 
\citet{FrebelAC05,FrebelJB07}, 
\citet{ThompsonIB08}, 
\citet{AokiBS08}, 
\citet{BeharaBL10}, 
\citet{PlaccoKB11}, 
\citet{CarolloBB12}, 
\citet{MasseronJP10,MasseronJL12}
\citet{YongNB13b}, and this paper. We did not considered the giants because their carbon abundance can be affected by mixing with deep layers 
\citep[see e.g., at low metallicity,][]{SpiteCP05,SpiteCH06}. In this figure we have kept the carbon abundance deduced from the CH molecular band and 1D computations to facilitate the comparison with the results in the literature since generally these previous abundances were derived from this feature and from 1D models. All stars in the graph are turnoff stars or dwarfs, the 3D correction must be about the same for all these stars, and, after this correction, the global view should remain similar.

According to the definition of the carbon-rich metal-poor stars ($\rm [C/Fe] > +1$) these stars are located above the dashed blue line in Fig. \ref{cdwarfs}. From this figure it seems that for $\rm [Fe/H] > -3$ the logarithm of the carbon abundance A(C) is almost constant close to 8.25, a value slightly lower than the solar value (A(C)=8.5).
This can reflect the fact that the carbon quantity transferred by the defunct AGB companion to the observed carbon-rich metal-poor star, has been such as to reach the same total amount in all stars, whatever the metallicity between --1.8 and --3 \citep[see][]{MasseronJP10}.

%$\bullet$ 
 The carbon abundance is clearly lower, with $\rm A(C)\approx 6.8$ for the low metallicities ($\rm [Fe/H] < -3.4$).
None of these stars have a (1D) carbon abundance significantly higher than A(C)=7.0.
However, the number of carbon-rich metal-poor stars with $\rm [Fe/H] < -3.4$ is very small and it is not possible to decide wether the mean value A(C)=6.8 represents a plateau of the carbon abundance or if it is an upper limit. Moreover, in this metallicity range, the stars are generally faint and a precise determination of the carbon abundance below A(C)=7.0 is difficult in turnoff stars. It would be interesting to find more CEMP dwarf stars with a temperature below 6000K and consequently a stronger CH absorption band, to reach a better determination of the behavior of the carbon abundance in the extremely metal-poor carbon stars.

%$\bullet$ 
In all stars with $\rm [Fe/H] < -3.4$, the barium abundance is also rather low: $\rm[Ba/Fe]<1.0$. All these stars belong to the class of the CEMP-no stars. Five other stars are located close to this plateau but with a metallicity higher than  $\rm [Fe/H] = -3.4$. In two of them (studied from high-resolution spectra) the abundance of Ba was measured. 
In BPS~CS\,29528-041, [Ba/Fe] is lower than 1.0. Therefore, this star belongs to the CEMP-no stars \citep{SivaraniBB06}, but SDSS\,J1036+1212 \citep{BeharaBL10} is Ba-rich ($\rm[Ba/Fe]=1.17$) and has a ratio $\rm[Eu/Fe]=1.26$. In this star $\rm[Ba/Eu]=-0.09$, this value is very close to $\rm[Ba/Eu]=0.0$ (within the error bar) and thus SDSS\,J1036+1212 is probably a CEMP-rs star \citep[see also the Fig.1 of][]{MasseronJP10}. 

%$\bullet$ 
The cause of the abundance anomalies of the CEMP-no stars are currently only poorly understood.   An enrichment in only C, N, and O by AGBs has been proposed \citep{MasseronJP10}, but in this case we would expect that many CEMP-no stars are binaries : only one CEMP-no star is known to be a binary \citep{PrestonS01},  and the radial velocity of SDSS\,J220924-002859 seems to be stable between 2003 and 2011 (see section \ref{radvel}).  However, more radial velocity measurements over a longer period of time would be necessary to exclude binarity. The abundance pattern of the CEMP-no stars could be also the result of a pre-enrichment of the interstellar medium  by e.g.  faint supernovae associated with the first generations of stars 
\citep{UmedaN03,UmedaN05,TominagaUN07}, or by C-rich winds of massive rotating EMP stars  \citep{HirschiFL06,MeynetEM06}.

To better interpret the abundance pattern of the CEMP-no stars, it is important to clearly establish wether the carbon abundance in this type of star is constant at a level of $\rm A(C)\approx 6.8$ or wether  this value is only an upper limit of the carbon abundance. It would be very important to significantly enlarge the sample of the abundance pattern of the very metal-poor CEMP stars and to study the scatter around the upper plateau, as well as around the possible lower plateau. Monitoring the radial velocities of the CEMP-no stars would also be important to research of the origin of the CEMP stars.

A 3D analysis of the CH and CN bands suggests that in metal-poor turnoff stars the carbon abundance could be 0.7 dex lower and the nitrogen abundance 2.4 dex lower than those derived in a 1D analysis of the CH G band or the CN band. This effect, which strongly depends on the external layers of the models, will be studied in detail in a forthcoming paper.

\begin {acknowledgements} 
The authors thanks the referee for his\textbackslash her very constructive remarks that helped us improve our manuscript.They thank David Yong for providing data in advance of publication
This work has been supported by the ``Programme National de Physique Stellaire'' and the ``Programme National de Cosmologie et Galaxies''  (CNRS-INSU).  EC, NC, and HGL acknowledge financial support by the Sonderforschungsbereich SFB 881 ``The Milky Way System (subprojects A4 and A5)'' of the German Research Foundation (DFG). 
\end {acknowledgements}

\bibliographystyle{aa}

\begin{thebibliography}{}

\bibitem[Abazajian et al.(2009)]{Abazajian09}
Abazajian K.N. et al., 2009 ApJS 182 543

\bibitem[Alvarez \& Plez(1998)]{AlvarezP98}
Alvarez R., Plez B., 1998, A\&A 330, 1109

\bibitem[Asplund(2005)]{Asplund05}
Asplund M., 2005, ARAA 43, 481

\bibitem[Andrievsky et al.(2007)]{AndrievskySK07}
%Na
Andrievsky S. M., Spite M., Korotin S. A., Spite F., Bonifacio P., AndrievskyAndrievskyAndrievskyCayrel R., Hill V., Fran\c cois P., 2007, A\&A 464, 1081

\bibitem[Andrievsky et al.(2008)]{AndrievskySK08}
%Al
Andrievsky S. M., Spite M., Korotin S. A., Spite F., Bonifacio P., Cayrel R., Hill V., Fran\c cois P., 2008, A\&A 481, 481
    
\bibitem[Andrievsky et al.(2009)]{AndrievskySK09}
%Ba 
Andrievsky S. M., Spite M., Korotin S.A., Spite F., Bonifacio P., Cayrel R., Hill V., Fran\c cois P.,  2009, A\&A 494, 1083

%\bibitem[Andrievsky et al.(2010)]{AndrievskySK10}
%Mg K 
%Andrievsky S. M., Spite M., Korotin S. A., Spite F., Bonifacio P., Cayrel R., Fran\c cois P., Hill V., 2010, A\&A 509, 88

\bibitem[Andrievsky et al.(2011)]{AndrievskySK11}
%Sr 
Andrievsky S. M., Spite F., Korotin S. A., Spite M. et al., 2011, A\&A 530, A105

\bibitem[Aoki et al.(2006)]{AokiFC06}
Aoki W., Frebel A., Christlieb N., Norris J. E., Beers T.C. et al., 2006, ApJ 639, 897

\bibitem[Aoki et al.(2009)]{AokiBS08}
Aoki W., Beers T.C., Sivarani T., Marsteller B., Lee Y.S., Honda S., Norris J.E., Ryan S.G., Carollo D., 2008, Apj 678, 1351

\bibitem[Barbuy et al.(2005)]{BarbuySS05}
Barbuy B., Spite M., Spite F., Hill V., Cayrel R., Plez B., Petitjean P., 2005, A\&A  429, 1031

\bibitem[Barklem et al.(2012)]{BarklemBS12} 
Barklem P.S., Belyaev A.K., Guitou M., Feautrier N., 2012, A\&A 541, A80

\bibitem[Beers \& Christlieb(2005)]{BeersC05}
Beers T.C., Christlieb N., 2005, ARA\&A 43,531

\bibitem[Behara et al.(2010)]{BeharaBL10}
Behara N. T., Bonifacio P., Ludwig H., et al. 2010, A\&A, 513, A72

\bibitem[Bergemann \& Cescutti(2010)]{BergemannC10}
Bergemann M., Cescutti G., 2010, A\&A 522, A9

\bibitem[Bergemann \& Gehren(2008)]{BergemannG08}
Bergemann M., Gehren T., 2008, A\&A 492, 823

\bibitem[Bisterzo et al.(2006)]{BisterzoGS06}
Bisterzo S., Gallino R., Straniero O., Ivans I.I., K\"appeler F., Aoki W., 2006, Mem S.A.It. 77, 985

\bibitem[Bonifacio et al.(2007)]{BonifacioMS07}
Bonifacio P., Molaro P., Sivarani T., Cayrel R., Spite M., Spite F., Plez B., Andersen J., Barbuy B., Beers T.C. et al., 2007, A\&A 462, 851

\bibitem[Bonifacio et al.(2009)]{BonifacioSC09} 
Bonifacio P., Spite M., Cayrel R., Hill V., et al.,  2009,
A\&A 501, 519 

%\bibitem[Bromm \& Loeb(2003)]{BrommL03}
%Bromm V, Loeb A., 2003, Nature 425, 812

%\bibitem[Bromm \& Larson, 2004]{BrommL04}
%Bromm V., Larson R.B., 2004, ARA\&A 42, 79

\bibitem[Caffau \& Ludwig(2007)]{CaffauL07}
Caffau E., Ludwig H.-G., 2007, A\&A 467, L11

\bibitem[Caffau et al.(2008)]{caffau08} 
Caffau, E., Ludwig, H.-G., Steffen, M., Ayres, T.~R., Bonifacio, P., Cayrel, R., Freytag, B., \& Plez, B.\ 2008, \aap, 488, 1031

\bibitem[Caffau et al.(2011)]{CaffauBF11}
Caffau E., Bonifacio P., Fran\c cois P., Sbordone L., Monaco L., Spite M., Spite F., Ludwig H.-G., Cayrel R., Zaggia S., et al., 2011, Nature 477, 67

\bibitem[Caffau et al.(2011)]{CaffauBFXh11}
Caffau E., Bonifacio P., Fran\c cois P., Spite M., Spite F., Zaggia S., Ludwig H.-G., Monaco L.,
Sbordone L., Cayrel R., Hammer F., Randich S., Hill V., Molaro P.,  2011, A\&A 534, A4

\bibitem[Caffau et al.,(2011)]{CaffauLS11} 
Caffau, E., Ludwig, H.-G., Steffen, M., Freytag, B., \& Bonifacio, P.,
2011, {\it Solar Phys.}, 268, 255

\bibitem[Caffau et al.(2012)]{CaffauBF12}
Caffau E., Bonifacio P., Fran\c cois P., Spite M., Spite F., Zaggia S., Ludwig H.-G., Steffen M., Mashonkina L., Monaco L., et al., 2012, A\&A 542, 51

\bibitem[Carollo et al.,(2011)]{CarolloBB12}
Carollo D., Beers T.C., Bovy J., Sivarani T., Norris J.E., Freeman K.C., Aoki W., Lee Y.S., Kennedy C.R., 2012, ApJ 744, 195

\bibitem[Cayrel(1988)]{Cayrel88}
Cayrel R., 1988, Proceedings of the IAU Symp 132 ``The impact of Very High S/N Spectroscopy on Stellar Physics'', G. Cayrel de Strobel and M.Spite eds, Kluwer Academic Publishers, P.345

\bibitem[Cayrel et al.,(2004)]{CayrelDS04}
Cayrel R., Depagne E., Spite M., Hill V., Spite F., Fran\c cois P., Plez B., Beers T.C., Primas F., Andersen J., Barbuy B., Bonifacio P.,Molaro P., Nordstr\"om B.,  2004 A\&A 416 1117 

\bibitem[Cescutti \& Chiappini(2013)]{CescuttiChiap13}
Cescutti G., Chiappini C., 2013, Proceedings of the XII International symposium on Nuclei in the Cosmos,  eds J. Lattanzio, A. Karakas, M. Lugaro, G. Dracoulis, http://pos.sissa.it, ArXiv:1301.1908

\bibitem[Christlieb et al.,(2002)]{ChristliebBB02}
Christlieb N., Bessell M. S., Beers T. C., Gustafsson B., Korn A., Barklem P. S., Karlsson T., Mizuno-Wiedner M., Rossi S., 2002, Nature 419, 904

\bibitem[Cohen et al.,(2003)]{CohenCQ03}
Cohen J.G., Christlieb N., Qian Y.-Z, Wasseburg G.J. et al., 2003, ApJ 588, 1082

\bibitem[Collet et al., 2007]{ColletAT07}
Collet, R., Asplund, M., \& Trampedach, R.
2007, A\&A, 469, 687

\bibitem[Dekker et al.(2000)]{DekkerDK00}
Dekker H., D'Odorico S., Kaufer A., Delabre B., Kotzlowski H., 2000, Proc. SPIE 4008, 534

\bibitem[Frebel et al.(2005)]{FrebelAC05}
Frebel A., Aoki W., Christlieb N., Ando H., Asplund M., Barklem P.S., Beers T.C., Eriksson K., Fechner C., Fujimoto M.Y., et al. 2005, Nature 434, 871

\bibitem[Frebel et al.(2006)]{FrebelCN06}
Frebel A., Christlieb N., Norris J.E., Beers T.C., Bessell M., J. Rhee et al., 2006, ApJ 652, 1585Beers T.C.,

\bibitem[Frebel et al.(2007)]{FrebelJB07}
Frebel A., Johnson J.L., Bromm V., 2007, MNRAS 380, L40

\bibitem[Frebel(2010)]{Frebel10}
Frebel A.,  2010, AN 331, 474

\bibitem[Freytag et al.(2012)]{Freytag2012}
Freytag,~B., Steffen,~M., Ludwig,~H.-G., 
Wedemeyer-Bh\"om,~S., Schaffenberger,~W., \& Steiner,~O.
2012, J. Comp. Phys., 231, 919

\bibitem[Gehren et al.(2004)]{GehrenLS04}
Gehren T., Liang Y.C., Shi J.R.et al.\ 2004, A\&A 413, 1045

\bibitem[Gonz\'alez Hern\'andez et al.(2008)]{GonzalezBL08}
Gonz\'alez Hern\'andez J.I., Bonifacio P., Ludwig H.-G., Caffau E., Spite M., Spite F., Cayrel R., Molaro P., Hill V., Fran\c cois P. et al., 2008, A\&A 480, 233

\bibitem[Gonz\'alez Hern{\'a}ndez et al., 2010]{iJGH10} 
Gonz{\'a}lez Hern{\'a}ndez, J.~I., 
Bonifacio, P., Ludwig, H.-G., Caffau, E., 
Behara, N.~T., \& Freytag, B.\ 
2010, A\&A, 519, A46 

%\bibitem[Gratton et al. 1999]{GrattonCE99}
%Gratton R.G., Carretta E., Eriksson K., Gustafsson B., 1999, A\&A, 350, 955

\bibitem[Gustafsson et al.(1975)]{GustafssonBE75}
Gustafsson B., Bell R. A., Eriksson K., Nordlund \AA., 1975, A\&A, 42, 407 

\bibitem[Gustafsson et al.(2003)]{GustafssonEE03}
Gustafsson B., Edvardsson B., Eriksson K., et al. 2003, in Stellar Atmosphere
Modeling, ed. I. Hubeny, D. Mihalas, \& K. Werner, ASP Conf. Ser., 288, 331 

\bibitem[Gustafsson et al.(2008)]{GustafssonEE08}
Gustafsson B., Edvardsson B., Eriksson K., Graae-J\o rgensen U., Nordlund
\AA., Plez B., 2008, A\&A, 486, 951

\bibitem[Herwig et al.(2004)]{Herwig04}
Herwig F., 2004, ApJS, 155, 651

\bibitem[Hill et al.(2002)]{HillPC02}
Hill V., Plez B., Cayrel R., Beers T.C., Nordstr\"om B., Andersen J., Spite M., Spite F., 
Barbuy B.,  Bonifacio P., Depagne E., Fran\c cois P., Primas F., 2002, A\&A, 387, 560


\bibitem[Hirschi et al., 2006]{HirschiFL06}
Hirschi R., Fr\"ohlich C., Liebend\"orfer M., Thielemann F.-K.,  RvMA 19, proceedings for 79th Annual Scientific Meeting of the Deutsche Astronomische Gesellschaft 2005, 101 (astro-ph/0601502)  

\bibitem[Karakas \& Lattanzio(2007)]{KarakasL07}
Karakas A., Lattanzio J.C., 2007, PASA 24, 103

\bibitem[Lai et al.(2008)]{LaiBJ08}
Lai D.K., Bolte M., Johnson J.A., Lucatello S. Heger A. Woosley S.E., 2008, ApJ 681, 1524

\bibitem[Lodders et al.(2009)]{LoddersPG09}
Lodders K., Palme H., Gail H.-P., 2009, in Landolt-B\"ornstein, New Series,  Volume VI/4B Chapter 4.4. Edited by J.E. Tr\"umper, Berlin Heifelberg New York: Springer-Verlag p. 560

\bibitem[Lucatello et al.(2005)]{LucatelloTB05}
Lucatello S., Tsangarides S., Beers T.C., Carretta E., Gratton R., Ryan S.G., 2005, ApJ 625, 825
 
\bibitem[Lucatello et al.(2006)]{LucatelloBC06}
Lucatello S., Beers, T. C., Christlieb, N., Barklem P.S., Rossi S., Marsteller B., Sivarani T., Lee Y.S., 2006, ApJ 652, L37

\bibitem[Ludwig(1992)]{LudwigPhD}
Ludwig H.-G., 1992, University of Kiel, PhD thesis

\bibitem[Ludwig et al.(2008)]{LudwigBC08}
Ludwig H.-G., Bonifacio P., Caffau E., Behara N. T., Gonz\'alez Hern\'andez J. I., Sbordone L., 2008, Physica Scripta Volume T, 133, 014037

\bibitem[Ludwig et al.(2009)]{Ludwig-cifist09} Ludwig, H.-G., Caffau, E., Steffen, M., Freytag, B., Bonifacio, P., \& Ku{\v c}inskas, A.\ 2009, MmSAI, 80, 711 

\bibitem[Marsteller et al.(2005)]{MarstellerBR05}
Marsteller B., Beers T. C., Rossi S., Christlieb N., Bessell M., Rhee J.,
2005, Nucl. Phys. A, 758, 312

\bibitem[Mashonkina et al.(1999)]{MashonkinaGB99}
Mashonkina L., Gehren T., Bikmaev I., 1999, A\&A 343, 519

\bibitem[Mashonkina et al.(2012)]{MashonkinaRF12}
Mashonkina L., Ryabtsev A., Frebel A., 2012, A\&A 540, A98

\bibitem[Mashonkina (2013)]{Mashonkina-Mg13}
Mashonkina L., 2013, A\&A (in press, arXiv:1212.3192) 

\bibitem[Masseron et al.(2010)]{MasseronJP10}
Masseron T., Johnson J.A., Plez B., Van Eck S., Primas F., Goriely S., Jorissen A., 2010, A\&A 509, A93

\bibitem[Masseron et al.(2012)]{MasseronJL12}
Masseron T., Johnson J.A., Lucatello S., Karakas A., Plez B., Beers T.C., Christlieb N., 2012, ApJ 751: 14 

\bibitem[Meynet et al.(2006)]{MeynetEM06}
Meynet G., Ekstr\"om S., Maeder A., 2006, A\&A 447, 623

\bibitem[Nordlund(1983)]{Nordlund83}
Nordlund, A., 1983, in Solar and stellar magnetic fields: Origins and coronal effects; Proceedings of the Symposium, Zurich, Switzerland, August 2-6, 1982 (A84-42426 20-90). Dordrecht, D. Reidel Publishing Co., p. 79.

\bibitem[Paunzen et al.(1999)]{Paunzen99} 
Paunzen, E., et al.\ 
1999, \aap, 345, 597 

\bibitem[Placco et al.(2011)]{PlaccoKB11}
Placco V.M., Kennedy C.R., Beers T.C., Christlieb N., Rossi S., Sivarani T., Lee Y.S., Reimers D., and Wisotzki L. 2011, ApJ 142, 188


\bibitem[Plez \& Cohen,(2005)]{PlezCo05}
Plez B., Cohen  J.~G, 2005, A\&A 434, 1117

\bibitem[Plez et al.(2008)]{Plez08}
Plez, B., Masseron, T., Van Eck, S., Jorissen, A., Coheur, P.F., Godefroid, M.,
Christlieb, N., 2008,   in 14th Cambridge Workshop on Cool Star, 
Stellar Systems, and the Sun, ed. G. van Belle, ASP Conf Series 384, CD-rom

\bibitem[Plez(2012)]{Plez12}
Plez B., 2012, ascl.soft05004P, http://adsabs.harvard.edu/abs/2012ascl.soft05004P

\bibitem[Preston \& Sneden(2001)]{PrestonS01}
Preston G. W., Sneden C., 2001, AJ 122, 1545

\bibitem[Rossi et al.(1999)]{RossiBS99}
Rossi S., Beers T. C., \& Sneden C. 1999, in ASP Conf. Ser. 165, The Third Stromlo Symposium: The Galactic Halo, eds. B. K. Gibson, R. S. Axelrod, \&
M. E. Putman (San Francisco: ASP), 264

%\bibitem[Santoro \& Shull(2006)]{SantoroS06}
%Santoro F., Shull J.M., 2006, ApJ 643, 26

\bibitem[Sbordone et al.(2010)]{SbordoneBC10}
Sbordone L., Bonifacio P., Caffau E., Ludwig H.-G., Behara N.T., Gonz\'alez Hern\'andez J.I., Steffen M., Cayrel R., Freytag B., van' t Veer C. et al., 2010, A\&A 522, 26

\bibitem[Schlegel et al.(1998)]{SchlegelFD98}
Schlegel D. J., Finkbeiner D. P., Davis M., 1998, ApJ 500, 525

%\bibitem[Schneider et al.(2006)]{SchneiderOI06}
%Schneider R.; Omukai K., Inoue A. K., Ferrara A., 2006, MNRAS, 369, 1437

\bibitem[Sivarani et al.(2006)]{SivaraniBB06}
Sivarani T., Beers T. C., Bonifacio P., Molaro P., Cayrel R., Herwig F., Spite M., Spite F., Plez B., Andersen J. et al., 2006, A\&A 459, 125

\bibitem[Spite et al.(2005)]{SpiteCP05}
Spite M., Cayrel R., Plez B., et al. 2005, A\&A, 430, 655

\bibitem[Spite et al.(2006)]{SpiteCH06}
Spite M., Cayrel R., Hill V., Spite F., Fran\c cois P., et al., 2006, A\&A, 455, 291

\bibitem[Spite et al.(2012)]{SpiteAS12}
Spite M., Andrievsky S. M., Spite F., Caffau E., Korotin S. A., Bonifacio P., Ludwig H.-G., Fran\c cois P.,  Cayrel R., 2012, 
A\&A 541, A143

\bibitem[Spite et al.(2012)]{SpiteSB12}
Spite M., Spite F., Bonifacio P., 2012, Mem. S.A.It. Suppl. Vol. 22, in press

\bibitem[Steenbock \& Holweger(1984)]{steen84} 
Steenbock, W., \& Holweger, H.\ 1984, \aap, 130, 319 

\bibitem[St\"urenburg \& Holweger(1990)]{SturH90}
St\"urenburg S., Holweger H., A\&A, 237, 125

\bibitem[Thompson et al.(2008)]{ThompsonIB08}
Thompson I.B., Ivans I.I., Bisterzo S., Sneden C., Gallino R., Vauclair S., Burley G.S., Shectman S.A., Preston G.W., 2008, ApJ 677, 556

\bibitem[Tominaga et al.(2007)]{TominagaUN07}
Tominaga N., Umeda H., Nomoto K., 2007, ApJ, 660, 516 

\bibitem[Umeda et al.(2003)]{UmedaN03}
Umeda H., \& Nomoto K., 2003, Nature, 422, 871

\bibitem[Umeda et al.(2005)]{UmedaN05}
Umeda H., \& Nomoto K., 2005, ApJ, 619, 427

\bibitem[Yong et al.(2013b)]{YongNB13b}
Yong D., Norris J.E., Bessell M.S., Christlieb N., Asplund M., Beers T.C., Barklem P.S., Frebel A., Ryan S.G., 2013, ApJ, 762:26 

\bibitem[Yong et al.(2013a)]{YongNB13a}
Yong D., Norris J.E., Bessell M.S., Christlieb N., Asplund M., Beers T.C., Barklem P.S., Frebel A., Ryan S.G., 2013, ApJ, 762:27 

\bibitem[York et al.(2000)]{Yorksdss00}
York D. G., et al. 2000, AJ, 120, 1579

\end{thebibliography}
{}
\onecolumn
\begin{longtab}
\begin{longtable}{|lcrr|rrr|rrr|rrr|}
\caption{\label{lines} Linelist equivalent widths and abundances}\\
\hline\hline
     &         &        &      &\multicolumn{3}{|c|}{SDSS\,J111407+182831}&\multicolumn{3}{|c|}{SDSS\,J114323+202058}&\multicolumn{3}{|c|}{SDSS\,J220924-002859}\\
Elem &$\lambda$&Exc.Pot.&\loggf&  EW  & abund & abund&  EW  & abund & abund&  EW  & abund & abund          \\
     & (\AA)   & (eV)   &      &(m\AA)&  LTE  & NLTE &(m\AA)&  LTE  & NLTE &(m\AA)&  LTE  & NLTE           \\  
\hline
\endfirsthead
\caption{continued.}\\
\hline\hline
     &         &        &      &\multicolumn{3}{|c|}{SDSS\,J111407+182831}&\multicolumn{3}{|c|}{SDSS\,J114323+202058}&\multicolumn{3}{|c|}{SDSS\,J220924-002859}\\
Elem &$\lambda$&Exc.Pot.&\loggf&  EW  & abund & abund&  EW  & abund & abund&  EW  & abund & abund          \\
     & (\AA)   & (eV)   &      &(m\AA)&  LTE  & NLTE &(m\AA)&  LTE  & NLTE &(m\AA)&  LTE  & NLTE           \\  
\hline
\endhead
 C  1& 4932.049&  7.69& -1.884&    syn&   8.40&  7.97&      syn&   8.20&  7.77&         &       &          \\
     &         &      &       &       &       &      &         &       &      &         &       &          \\
 O  1& 7771.941&  9.15&  0.369&    syn&       &      &      syn&   7.20&  7.07&      syn&   7.10&  6.97    \\
 O  1& 7774.161&  9.15&  0.223&    syn&  7.45 & 7.31 &      syn&   6.90&  6.82&      syn&   7.10&  6.97    \\
 O  1& 7775.388&  9.15&  0.001&    syn&       &      &         &       &      &         &       &          \\
     &         &      &       &       &       &      &         &       &      &         &       &          \\
Na  1& 5889.951&  0.00&  0.117&  149.8&   4.67&  4.57&    142.6&   4.63&  4.53&         &       &          \\
Na  1& 5895.924&  0.00& -0.184&  127.2&   4.69&  4.59&    122.0&   4.66&  4.56&         &       &          \\
     &         &      &       &       &       &      &         &       &      &         &       &          \\
Mg  1& 3829.355&  2.71& -0.231&    syn&   5.35&  5.49&      syn&   4.90&  5.00&      syn&   3.65&  3.90    \\
Mg  1& 3832.304&  2.71&  0.146&    syn&   5.50&  5.64&      syn&   5.05&  5.15&      syn&   3.95&  4.20    \\
Mg  1& 3838.290&  2.72&  0.415&    syn&   5.55&  5.69&      syn&   5.05&  5.15&      syn&   3.95&  4.20    \\
Mg  1& 4702.991&  4.35& -0.666&   39.3&   5.64&  5.78&     14.5&   5.11&  5.21&         &       &          \\
     &         &      &       &       &       &      &         &       &      &         &       &          \\
Al  1& 3961.520&  0.01& -0.323&   48.0&   2.83&  3.48&     31.5&   2.55&  3.20&         &       &          \\
     &         &      &       &       &       &      &         &       &      &         &       &          \\
Ca  1& 4226.728&  0.00&  0.244&   81.2&   3.30&  3.42&     87.8&   3.49&  3.64&     42.1&   2.60&  2.82    \\
Ca  1& 4454.779&  1.90&  0.258&   17.2&   3.57&  3.70&     16.7&   3.54&  3.70&         &       &          \\
Ca  1& 6162.173&  1.90& -0.090&    5.4&   3.25&  3.43&     13.6&   3.71&  3.89&         &       &          \\
Ca  2& 3933.680&  0.00&  0.105&    syn&   3.45&  3.44&      syn&   3.55&  3.55&      syn&   2.85&  2.82    \\
Ca  2& 8498.023&  1.69& -1.416&  118.2&   3.65&  3.30&    133.0&   3.87&  3.55&     92.6&   3.35&  3.05    \\
Ca  2& 8542.091&  1.70& -0.463&  206.4&   3.50&  2.99&    250.0&   3.74&  3.19&    180.6&   3.47&  2.87    \\
Ca  2& 8662.170&  1.69& -0.723&  211.3&   3.76&  3.24&    215.0&   3.82&  3.29&    134.5&   3.36&  2.81    \\
     &         &      &       &       &       &      &         &       &      &         &       &          \\
Sc  2& 4246.822&  0.31&  0.242&   47.2&   0.71&      &     25.7&   0.27&      &         &       &          \\
     &         &      &       &       &       &      &         &       &      &         &       &          \\
Ti  1& 4981.731&  0.85&  0.504&    syn&   2.1 &      &         &       &      &         &       &          \\
Ti  2& 3759.291&  0.61&  0.280&    syn&   1.7 &      &     70.4&   2.03&      &         &       &          \\
Ti  2& 3761.320&  0.57&  0.180&    syn&   1.7 &      &     67.4&   2.02&      &         &       &          \\
Ti  2& 3913.461&  1.12& -0.420&   16.0&   1.81&      &     20.0&   1.95&      &         &       &          \\
Ti  2& 4417.714&  1.16& -1.190&    syn&   2.1 &      &      5.2&   2.05&      &         &       &          \\
Ti  2& 4443.794&  1.08& -0.720&   12.4&   1.90&      &     17.1&   2.08&      &         &       &          \\
Ti  2& 4450.482&  1.08& -1.520&    syn&   2.0 &      &      5.1&   2.29&      &         &       &          \\
Ti  2& 4464.449&  1.16& -1.810&       &       &      &      4.3&   2.57&      &         &       &          \\
Ti  2& 4468.507&  1.13& -0.600&   19.1&   2.05&      &     14.7&   1.93&      &         &       &          \\
Ti  2& 4501.270&  1.12& -0.770&   10.3&   1.88&      &     10.1&   1.89&      &         &       &          \\
Ti  2& 4533.960&  1.24& -0.530&    syn&   1.75&      &     19.1&   2.09&      &         &       &          \\
Ti  2& 4563.757&  1.22& -0.690&    8.4&   1.80&      &      8.1&   1.80&      &         &       &          \\
Ti  2& 4571.968&  1.57& -0.320&    syn&   1.8 &      &     20.4&   2.23&      &         &       &          \\
     &         &      &       &       &       &      &         &       &      &         &       &          \\
V   1& 4379.230&  0.30&  0.580&   10.8&   2.25&      &         &       &      &         &       &          \\
     &         &      &       &       &       &      &         &       &      &         &       &          \\
Cr  1& 4254.336&  0.00& -0.114&    syn&   2.05&  2.49&     22.1&   2.25&  2.69&         &       &          \\
     &         &      &       &       &       &      &         &       &      &         &       &          \\
Mn  1& 4030.753&  0.00& -0.470&       &       &  2.00&      syn&   1.95&      &         &       &          \\
Mn  1& 4033.062&  0.00& -0.618&    syn&   1.50&  2.00&      syn&   1.85&      &         &       &          \\
     &         &      &       &       &       &      &         &       &      &         &       &          \\
Fe  1& 3758.233&  0.96& -0.027&   69.0&   4.05&      &     78.8&   4.37&      &         &       &          \\
Fe  1& 3763.789&  0.99& -0.238&   51.9&   3.80&      &     72.2&   4.42&      &         &       &          \\
Fe  1& 3767.192&  1.01& -0.389&   54.2&   4.03&      &     59.7&   4.22&      &         &       &          \\
Fe  1& 3786.677&  1.01& -2.225&    7.6&   4.51&      &      8.1&   4.58&      &         &       &          \\
Fe  1& 3787.880&  1.01& -0.859&   35.4&   4.04&      &     45.5&   4.31&      &         &       &          \\
Fe  1& 3815.840&  1.49&  0.237&   61.6&   4.03&      &     73.5&   4.40&      &         &       &          \\
Fe  1& 3820.425&  0.86&  0.119&   86.2&   4.26&      &     93.3&   4.48&      &         &       &          \\
Fe  1& 3824.444&  0.00& -1.362&   62.1&   4.29&      &     66.0&   4.45&      &         &       &          \\
Fe  1& 3827.822&  1.56&  0.062&   52.6&   4.03&      &     54.4&   4.11&      &         &       &          \\
Fe  1& 3840.437&  0.99& -0.506&   54.3&   4.12&      &     67.2&   4.51&      &         &       &          \\
Fe  1& 3849.966&  1.01& -0.871&   46.4&   4.30&      &     53.6&   4.52&      &         &       &          \\
Fe  1& 3850.818&  0.99& -1.734&   16.7&   4.40&      &     20.4&   4.55&      &         &       &          \\
Fe  1& 3856.371&  0.05& -1.286&   63.3&   4.29&      &     67.3&   4.46&      &         &       &          \\
Fe  1& 3859.911&  0.00& -0.710&   75.3&   4.04&      &     81.7&   4.28&      &         &       &          \\
Fe  1& 3865.523&  1.01& -0.982&   34.0&   4.12&      &     54.6&   4.65&      &         &       &          \\
Fe  1& 3878.018&  0.96& -0.914&   45.8&   4.27&      &     55.5&   4.56&      &         &       &          \\
Fe  1& 3899.707&  0.09& -1.531&   55.1&   4.32&      &     58.5&   4.46&      &         &       &          \\
Fe  1& 3920.258&  0.12& -1.746&   38.2&   4.14&      &     49.0&   4.44&      &         &       &          \\
Fe  1& 3922.912&  0.05& -1.651&   43.7&   4.11&      &     50.4&   4.32&      &         &       &          \\
Fe  1& 4005.242&  1.56& -0.610&   29.3&   4.13&      &     33.8&   4.27&      &         &       &          \\
Fe  1& 4045.812&  1.49&  0.280&   64.6&   4.03&      &     69.1&   4.19&      &     36.4&   3.52&          \\
Fe  1& 4063.594&  1.56&  0.062&   60.8&   4.20&      &     64.2&   4.33&      &     31.7&   3.69&          \\
Fe  1& 4071.738&  1.61& -0.022&   46.9&   3.98&      &     57.7&   4.29&      &     28.9&   3.75&          \\
Fe  1& 4076.629&  3.21& -0.529&   10.5&   4.97&      &         &       &      &         &       &          \\
Fe  1& 4132.058&  1.61& -0.675&   28.1&   4.20&      &     24.8&   4.15&      &         &       &          \\
Fe  1& 4143.868&  1.56& -0.511&   35.1&   4.15&      &     33.0&   4.13&      &         &       &          \\
Fe  1& 4157.780&  3.42& -0.403&   26.6&   5.54&      &     14.9&   5.22&      &         &       &          \\
Fe  1& 4199.095&  3.05&  0.155&   44.6&   5.07&      &     39.9&   4.99&      &         &       &          \\
Fe  1& 4202.029&  1.49& -0.708&   24.5&   4.02&      &     34.8&   4.30&      &         &       &          \\
Fe  1& 4222.213&  2.45& -0.967&       &       &      &      9.5&   4.68&      &         &       &          \\
Fe  1& 4227.427&  3.33&  0.266&   13.9&   4.41&      &     17.4&   4.56&      &         &       &          \\
Fe  1& 4250.119&  2.47& -0.405&   16.1&   4.37&      &     11.0&   4.20&      &         &       &          \\
Fe  1& 4260.474&  2.40&  0.109&   26.5&   4.09&      &     26.5&   4.11&      &         &       &          \\
Fe  1& 4383.545&  1.49&  0.200&   64.4&   4.05&      &     67.6&   4.18&      &     28.8&   3.39&          \\
Fe  1& 4404.750&  1.56& -0.142&   41.9&   3.91&      &     52.3&   4.18&      &     17.4&   3.49&          \\
Fe  1& 4443.194&  2.86& -1.043&    5.1&   4.79&      &      5.2&   4.82&      &         &       &          \\
Fe  1& 4466.551&  2.83& -0.600&    7.4&   4.50&      &      7.5&   4.53&      &         &       &          \\
Fe  1& 4489.739&  0.12& -3.966&    8.6&   5.41&      &      4.9&   5.19&      &         &       &          \\
Fe  1& 4528.614&  2.18& -0.822&   10.4&   4.28&      &     13.8&   4.45&      &         &       &          \\
Fe  1& 4531.148&  1.49& -2.155&    5.5&   4.66&      &      4.3&   4.58&      &         &       &          \\
Fe  1& 4891.492&  2.85& -0.112&   16.5&   4.40&      &         &       &      &         &       &          \\
Fe  1& 4918.994&  2.87& -0.342&       &       &      &     11.7&   4.49&      &         &       &          \\
Fe  1& 4920.502&  2.83&  0.068&   11.3&   4.01&      &     18.5&   4.29&      &         &       &          \\
Fe  2& 4233.172&  2.58& -1.947&    8.2&   4.23&      &      7.7&   4.20&      &         &       &          \\
Fe  2& 4583.837&  2.81& -1.867&    5.3&   4.12&      &      8.7&   4.37&      &         &       &          \\
Fe  2& 4923.927&  2.89& -1.320&   16.0&   4.18&      &     15.0&   4.15&      &         &       &          \\
     &         &      &       &       &       &      &         &       &      &         &       &          \\
Co  1& 3845.461&  0.92&  0.010&   16.5&   2.46&      &      syn&   2.3 &      &         &       &          \\
Co  1& 3995.302&  0.92& -0.220&   10.0&   2.41&      &      syn&   2.2 &      &         &       &          \\
Co  1& 4121.311&  0.92& -0.320&    7.3&   2.34&      &      6.0&   2.29&      &         &       &          \\
     &         &      &       &       &       &      &         &       &      &         &       &          \\
Ni  1& 3807.138&  0.42& -1.205&   15.1&   2.96&      &      syn&   3.12&      &         &       &          \\
Ni  1& 3858.292&  0.42& -0.936&   23.8&   2.95&      &      syn&   2.85&      &         &       &          \\
     &         &      &       &       &       &      &         &       &      &         &       &          \\
Zn  1& 4810.528&  4.08& -0.137&    2.5&   2.09&      &      2.7&   2.14&      &         &       &          \\
     &         &      &       &       &       &      &         &       &      &         &       &          \\
Sr  2& 4077.709&  0.00&  0.167&   64.4&   0.11&  0.31&     70.9&   0.34&  0.54&  $<8.0$ &$<-1.45$&$<-1.20$ \\
Sr  2& 4215.519&  0.00& -0.145&   53.5&   0.06&  0.26&     59.3&   0.26&  0.46&         &        &         \\
     &         &      &       &       &       &      &         &       &      &         &        &         \\
Ba  2& 4554.029&  0.00&  0.170&    syn&   0.41&  0.41&      syn&   0.95&  0.95& $<10.0$ &$<-1.04$&$<-1.04$ \\
Ba  2& 5853.668&  0.60& -1.000&       &       &      &      syn&   0.80&  0.80&         &        &         \\
Ba  2& 6141.713&  0.70& -0.076&   36.6&   0.42&  0.42&     52.5&   0.85&  0.85&         &        &         \\
Ba  2& 6496.897&  0.60& -0.377&   32.0&   0.50&  0.50&     44.6&   0.83&  0.83&         &        &         \\
\hline
\end{longtable}
\end{longtab}
\end{document}